\documentclass[aps,prd,superscriptaddress,showkeys,showpacs,twocolumn,a4paper]{revtex4-1}
\usepackage{ulem}
\usepackage{color}
\usepackage{graphicx}
\usepackage{soul}
\usepackage{appendix}
\usepackage{epsfig}
\usepackage[usenames,dvipsnames,svgnames]{xcolor}
\definecolor{purple}{rgb}{0.3,0,0.9} 
\definecolor{darkteal}{HTML}{045D5D}
\usepackage[colorlinks=true,linkcolor= cyan, citecolor = blue, urlcolor = cyan ]{hyperref}
\usepackage{epstopdf}

\usepackage{amsmath,amssymb,amsthm,amsfonts} 
\usepackage{subfig}
\usepackage{comment}
\usepackage{float}

\usepackage{float}

\newcommand{\be}{\begin{equation}}
\newcommand{\ee}{\end{equation}}
\newcommand{\ba}{\begin{eqnarray}}
\newcommand{\ea}{\end{eqnarray}}
\newcommand{\nn}{\nonumber\\}

\begin{document}
\title{Study of the heavy quarks energy loss through medium polarization, elastic collision and radiative processes}
\author{Jai Prakash}
	\email{jaiprakashaggrawal2@gmail.com}
        \affiliation{Department of Physics, Indian Institute of Technology Bombay, Mumbai 400076, India}
        \affiliation{School of Physical Sciences, Indian Institute of Technology Goa, Ponda-403401, Goa, India}
\author{Mohammad Yousuf Jamal}
	\email{yousufjml5@gmail.com}
        \affiliation{School of Physical Sciences, Indian Institute of Technology Goa, Ponda-403401, Goa, India}

\begin{abstract}
Heavy quarks serve as crucial probes for exploring the properties of the hot and dense medium formed in heavy-ion collision experiments. Understanding the modification of their energy as they traverse the medium is a focal point of research, with various authors extensively studying this phenomenon. This study specifically concentrates on the equilibrium phase, the quark-gluon plasma, and offers a comparative analysis of heavy quark energy loss through medium polarization, elastic collisions, and radiation. Notably, while previous studies have compared polarization loss and radiation, our work extends this by incorporating elastic collisions for a more comprehensive examination. The significance of medium polarization, particularly at low momentum, is underscored, as it has been found to contribute substantially. The formalism for energy loss and drag coefficient in each case is presented, followed by the calculation of the nuclear modification factor ($R_{AA}$) for a holistic comparative study.
\end{abstract}
\maketitle

\section{Introduction}
\label{sec:intro}

The study of Quark-Gluon Plasma (QGP), a hot and dense state of matter that is believed to resemble the state of the universe shortly after the Big Bang, has been made possible through the examination of heavy-ion collisions (HICs)~\cite{PHOBOS:2004zne, BRAHMS:2004adc, ALICE:2010khr, Fukushima:2020yzx}. However, the QGP medium produced in the experiment is short-lived ( $\sim $ 1 fm to 10 fm), and it is not possible to send any external probe to investigate the medium. Therefore, internal, self-generated probes are relied upon. The heavy quarks (HQs), distinguished by their substantially higher mass ($M_{HQ} \gg T$, where $M_{HQ}$ is the mass of HQs, and $T$ denotes the QGP medium temperature), are primarily generated in the initial stages of collisions (formation time, $\tau \sim 1/M$) behave as independent degrees of freedom throughout the evolving medium and hence, prove to be effective self-generated probes~\cite{vanHees:2005wb, Das:2009vy, Das:2013kea, Cao:2018ews, Song:2020tfm, Kurian:2020orp, Sebastian:2022sga}. The HQs interact with the QGP medium also and lose energy through various mechanisms, which is a complex phenomenon that depends on several factors, including the HQ's mass, energy, and the properties of the medium.

 While several studies have investigated the energy loss of HQs through various mechanism separately~\cite{Singh:2015eta, Matsui:1986dk, Jamal:2018mog, Agotiya:2016bqr,Song:2015sfa, Rapp:2018qla,   GolamMustafa:1997id, Plumari:2017ntm, Gossiaux:2008jv, Prakash:2021lwt, Prakash:2023wbs, Jamal:2023ncn, Singh:2023smw, Cao:2016gvr, Mazumder:2011nj, Zhang:2022fum, PhysRevD.103.054030, Jamal:2021btg, Jamal:2020emj, Sun:2023adv, Plumari:2019hzp, du2023accelerated,Shaikh:2021lka,Kumar:2021goi,Das:2022lqh,Bandyopadhyay:2023hiv,Romatschke:2007mq, Ryu:2015vwa,Prakash:2024rdz,Das:2022lqh,Ruggieri:2022kxv,Sumit:2023oib,Du:2023ewh,Prakash:2024irm}, a comprehensive comparison between them is still needed to understand their relative contributions in the QGP medium.
 There are broadly two ways that contribute to the energy loss of the HQ while passing through the QGP medium -- soft and hard contributions. Soft contribution arises due to medium polarization, which causes a back reaction to the HQ from the chromoelectric field generated from the HQ's motion itself \cite{Carrington:2015xca, Carrington:2016mhd, Jamal:2020fxo, YousufJamal:2019pen, Chakraborty:2006db, GolamMustafa:1997id}. Hard contribution arises due to collisions with hard particles, which can be elastic or inelastic. Elastic collisions occur when the HQ scatters with the medium particles, $2 \rightarrow 2$, leading to the transfer of energy from the HQ to the medium. Inelastic collisions, on the other hand, involve the production of secondary particles, $2 \rightarrow 3$, that carry away energy from the HQ, such as radiation or gluon emission. In the equilibrated plasma, a significant portion of energy loss occurs due to interactions with medium particles. However, the soft contribution in which the HQ interacts with gauge fields or plasma modes also contributes significantly to its energy loss as it travels through the QGP medium, especially at low momentum \cite{Carrington:2015xca, Carrington:2016mhd, Jamal:2020fxo, YousufJamal:2019pen, Chakraborty:2006db, GolamMustafa:1997id}. In this case, the colored HQs traverse the equilibrated QGP medium, inducing a chromoelectric field. This field opposes the HQ's motion, leading to a back reaction that hinders its propagation. Therefore, the HQs must spend energy to overcome the induced field, resulting in energy loss. The measurement of this energy loss in the HQs is based on the work required to counter the retarding forces created by the induced chromoelectric field. The polarization effects of the medium can be studied to analyze this field, and the dynamics of the HQs can be described by the Wong equations while treating them as a classical particle within the $SU(N_c)$ gauge group. The Boltzmann transport equation can be used to obtain the induced field. Recently, polarization loss has been studied in the presence of chirality and magnetic fields \cite{Ghosh:2023ghi, K:2023dum}, as well as in the Gribov plasma \cite{ Debnath:2023zet}. Following our recent work, where we compared the polarization loss with inelastic collision (radiation) \cite{Prakash:2023zeu}, in this study, we aim to incorporate elastic collisions for completeness. We shall compare the contribution of elastic collisions, medium polarization, and radiation through the energy loss of HQs, drag coefficient, and $R_{AA}$.

The remaining part of this manuscript is organized as follows. In Sec.~\ref{GSE}, the formalisms for each approach are discussed. Section~ \ref{RD} presents a comparative study of the three processes and discusses the results. Section~ \ref{SC} summarizes the findings, draws conclusions, and outlines future directions for this analysis. Throughout this work, natural units are employed, and the metric tensor is defined as $g_{\mu\nu}={\text {diag}}(1,-1,-1,-1)$.

\section{Methodology}
\label{GSE}
As mentioned earlier, our current investigation aims to comprehensively examine the HQ energy loss through three different processes: medium polarization, elastic collisions, and inelastic collisions. We have already discussed the formalism of polarization and radiation in one of our previous articles ~\cite{Prakash:2023zeu}. Therefore, we will keep it brief and focus more on the elastic collision mechanisms. We will introduce distinct notations for each case, as needed, to ensure clarity and avoid any potential overlap in the analyses. This meticulous approach will allow us to gain a fine understanding of the complex dynamics associated with each energy loss process.

\subsection{Energy loss through elastic collisions}

In our analysis, we use the formalism established by Svetitsky~\cite{Svetitsky:1987gq}, which allows us to treat the evolution of the HQs in the medium as Brownian motion. The dynamics of the HQs are characterized by a distribution function framed within the context of transport theory as ~\cite{Walton:1999dy, Moore:2004tg},

\begin{equation}\label{11}
p^{\mu}\partial_{\mu}f_{HQs}=\bigg(\dfrac{\partial f_{HQs}}{\partial t}\bigg)_{\text{int}},
\end{equation}

where $f_{HQs}$ represents the momentum distribution and $\bf p$ ($p=|{\bf p}|$) is the momentum of the HQs. Next, we are assuming that the medium is uniform, hence ignoring all the background medium interactions. Under these assumptions, the modifications to this distribution function are due to the elastic and inelastic interactions of the HQ within the medium. These interactions impact on the rate of change of $f_{HQs}$ and can be expressed quantitatively by the collision term $\bigg(\dfrac{\partial f_{HQs}}{\partial t}\bigg)_{col}$ given as follow, 

\begin{align}\label{12}
   \bigg(\dfrac{\partial f_{HQs}}{\partial t}\bigg)_{col}=\int{d^3{\bf Q}\bigg[\omega({\bf p}+{\bf Q},{\bf Q})f_{HQs}({\bf p}+{\bf Q})}\nonumber\\
   -\omega({\bf p},{\bf Q})f_{HQs}({\bf p})\bigg],
\end{align}
where ${\bf Q}= {\bf p}-{\bf p'}$ is the momentum transfer in the collision and $\omega({\bf p, Q})$ is the collision rate of the HQs. After a collision, the HQ momentum is reduced to $\bf p-Q$. This formalism is further taken in the approximation of (${\bf |p|}\gg {\bf |Q|}$), where a HQ collides with the medium's light quarks, anti-quarks, and gluons. The gain in the probability of the HQs is described by $\omega({\bf p+Q,Q})f_{HQs}({\bf p,Q})$. This quantity can be expanded in the power of ${\bf Q}$  using Taylor series expansion as follows,

\begin{align}\label{13}
   \omega({\bf p+Q,Q})f_{HQs}({\bf p,Q})\approx&\,\omega({\bf p,Q})f_{HQs}({\bf p})+{\bf Q}.\frac{\partial}{\partial {\bf p}}(\omega f_{HQs})\nonumber\\ &+\frac{1}{2}Q_iQ_j\frac{\partial^2}{\partial p_i\partial p_j}(\omega f_{HQs}).
\end{align}
One can achieve the differential Fokker-Planck equation from the Boltzmann transport equation as,

\begin{align}\label{14}
  	\frac{\partial f_{HQs}}{\partial t}=\frac{\partial}{\partial p_i}\left[\gamma_i({\bf p})f_{HQs}+\frac{\partial}{\partial p_j}\Big(D_{i j}({\bf p})f_{HQs}\Big)\right],
  	\end{align}

where,   

  \begin{align}
 \gamma_i=\int d^3Q\omega(p,Q)Q_i,
 \end{align}
 and, 
 \begin{align}
 D_{ij}=\frac{1}{2}\int d^3Q\omega(p,Q)Q_iQ_j,
\end{align}
 rewriting $\gamma_i$ and $D_{ij}$ for the elastic collisions, we have, 
 \begin{align}\label{15}
    \gamma_i=&\frac{1}{2{\text E}}\int{\frac{d^3{\bf q}}{(2\pi)^32{\text E}_q}}\int{\frac{d^3{\bf q}'}{(2\pi)^32{\text E}_{q'}}}\int{\frac{d^3{\bf p}'}{(2\pi)^32{\text E}_{p'}}}\frac{1}{\gamma_{HQs}}\nonumber\\ &\times\sum|\mathcal{M}_{2\rightarrow 2}|^2(2\pi)^4\delta^4 (P+O-P'-O') f_{k}({\bf{q}})\nonumber\\ &\times\Big(1+a_k f_{k}({\bf{q'}})\Big)\Big[({\bf p}-{\bf p}')_i\Big]\nonumber\\
	&=\langle\langle({\bf p}-{\bf p}')_i\rangle\rangle,
\end{align}
 \begin{align}\label{15_D}
    D_{ij}=&\frac{1}{2{\text E}}\int{\frac{d^3{\bf q}}{(2\pi)^32{\text E}_q}}\int{\frac{d^3{\bf q}'}{(2\pi)^32{\text E}_{q'}}}\int{\frac{d^3{\bf p}'}{(2\pi)^32{\text E}_{p'}}}\frac{1}{\gamma_{HQs}}\nonumber\\ &\times\sum|\mathcal{M}_{2\rightarrow 2}|^2(2\pi)^4\delta^4 (P+O-P'-O') f_{k}({\bf{q}})\nonumber\\ &\times\Big(1+a_k f_{k}({\bf{q'}})\Big)\Big[({\bf p}-{\bf p}')_i({\bf p}-{\bf p}')_i\Big]\nonumber\\
	&=\langle\langle({\bf p}-{\bf p}')_i({\bf p}-{\bf p}')_j\rangle\rangle.
\end{align}
The scattering matrix element, $\sum|\mathcal{M}_{2\rightarrow 2}|^2$ \cite{Svetitsky:1987gq} upholds the information of the HQs interaction with the medium particles \cite{Svetitsky:1987gq}. Here, $P$ is the four-momentum of the HQs, and $O$ is the four-momentum of the medium particle before the collision. Let $P'$ be the four-momentum of the HQs, and $O'$ be the four-momentum of the medium particle after the collision. The energy, ${\text E}_q$, and the momentum, $\bf{q}$ of medium particle correspond to before collision whereas, ${\text E}_q'$, and $\bf{q'}$ indicate the after collision scenario. The delta function ensures energy-momentum conservation. The phase space distribution function of the medium particles, $f_{k}({\bf{q'}})$, gives the temperature dependence of the drag coefficient. Next, the momentum dependency of $\gamma$  can be written as follow \cite{Svetitsky:1987gq},

\begin{align}\label{17}
 &\gamma(p^2)=\langle\langle 1 \rangle\rangle - \frac{\langle\langle {\bf{p.p'} \rangle\rangle}}{p^2}.
\end{align}
Similarly, the momentum diffusion $D_{ij}$ can be decomposed in terms of longitudinal and transverse components as follows, 
\begin{align}\label{18}
&D_{i,j} = \left(\delta_{ij}-\frac{p_ip_j}{p^2}\right) D_0(p^2)+\frac{p_ip_j}{p^2}D_1(p^2),
\end{align}
where $D_0$ is the transverse  and $D_1$ is the longitudinal diffusion coefficients, and $D_0$ take the forms of,
\begin{align}\label{19}
&D_{0}= \frac{1}{4}\left[\langle\langle p'^{2} \rangle\rangle-\frac{\langle\langle ({\bf{p'.p}})^2\rangle\rangle}{p^2} \right].
\end{align}
the average of a function, $\langle \langle F(\textbf{p})\rangle\rangle$ for the collisional processes in a QCD medium as follows, 
 
\begin{align}\label{avergef}
   \langle \langle F(\textbf{p})\rangle\rangle&=\frac{1}{2{\text E}}\int{\frac{d^3{\bf q}}{(2\pi)^32{\text E}_q}}\int{\frac{d^3{\bf q}'}{(2\pi)^32{\text E}_{q'}}}\int{\frac{d^3{\bf p}'}{(2\pi)^32{\text E}_{p'}}}\nonumber\\ &\times\frac{1}{\gamma_{HQ}}\sum|\mathcal{M}_{2\rightarrow 2}|^2(2\pi)^4\delta^4 (P+O-P'-O') \nonumber\\ &\times f^0_{k}({\bf{q}})\Big(1+a_k f^0_{k}({\bf{q'}})\Big)F({\textbf{p}'}),
\end{align}
Fermi suppression corresponds to $a_k=-1$, and Bose enhancement corresponds to $a_k=+1$ for the final state phase space of the medium constituent. The Mandelstam variable $s$ represents the square of the center-of-mass energy. Infrared divergence in the gluonic propagator within the $t-$ channel diagrams is screened by the Debye mass. The transport coefficient of the HQ due to elastic scattering can be schematically written as:
\begin{align}\label{24}
    X_c=\int \text{phase space}\times \text{ interaction}\times \text{transport part}.
\end{align}
It is to be noted that we are avoiding the diffusion part $D_{i j}({\bf p})$ as it is beyond the scope of the present analysis and restricting this investigation to the drag effect by considering only the drag coefficient. Next, we shall discuss the energy loss due to inelastic collision or radiation.

\subsection{Energy loss through radiation}

The high-energy HQs can also emit gluons in the hot QGP medium as they move through it. The radiative process $2\rightarrow 3$ is considered as follows: $HQs(P)+l(O)\rightarrow HQs(P^{'})+l(O^{'})+g(K_5)$, where $K_5=(E_5, k_{\perp}, k_z)$ is the four-momentum of the emitted gluons from the HQs, and $E_5$ is the energy of the emitted gluon.  The  radiation  transport coefficient ($X_{r}$) for this process can be described as~\cite{Mazumder:2013oaa, Shaikh:2021lka, Prakash:2021lwt, Liu:2020dlt, Song:2022wil},

\begin{widetext}
\begin{align}\label{rad}
X_{r}&=\frac{1}{2{\text E}}\int{\frac{d^3{\bf q}}{(2\pi)^32{\text E}_q}}\int{\frac{d^3{\bf q}'}{(2\pi)^32{\text E}_{q'}}}\int{\frac{d^3{\bf p}'}{(2\pi)^32{\text E}_{p'}}} \int{\frac{d^3{\bf k}_5}{(2\pi)^32{\text E}_5}}\frac{1}{\gamma_{HQ}}\sum{|\mathcal{M}|^2_{2\rightarrow 3}(2\pi)^4\delta^4 }(P+Q-P'-Q'-K_5)\nonumber\\ 
&\times f_k^0({\bf q})(1+a_k f^0_{k}({\bf q}'){F({\textbf{p}'})}\Theta_1({\text E}-{\text E}_5)\Theta_2(\tau-\tau_F),
\end {align}
\end{widetext}

{ where the $\gamma_{HQ}$ denotes statistical degeneracy factor of the HQs}. This analysis takes soft gluon emission, with the momentum of the emitted gluon $k_5 \rightarrow 0$. The radiated gluons follow the Bose-Einstein distribution, $\hat{f}(E_5)=\frac{1}{\exp{(\beta E_5)}-1}$, where $\beta$ represents  the inverse temperature of the system. The emission of gluons from HQs in a medium is subject to two constraints. The first constraint, $\Theta_1(\text{E}-E_5)$, ensures $E_5$ $<<$ $\text{E}$, as the HQ's energy limits gluon emission energy. The second constraint, $\Theta_2(\tau-\tau_F)$, ensures the gluon formation time ($\tau_F$) is shorter than the HQ's scattering time($\tau$), accounting for the Landau-Pomeranchuk-Migdal effect~\cite{Gyulassy:1993hr, Klein:1998du}. The invariant amplitude for inelastic processes, $|\mathcal{M}|^2_{2\rightarrow 3}$, can be defined in terms of the elastic invariant amplitude $|\mathcal{M}|^2_{2\rightarrow 2}$ \cite{Mazumder:2015wma,Abir:2011jb},

\begin{align}\label{matrixr}
    |\mathcal{M}|^2_{2\rightarrow 3}=|\mathcal{M}|^2_{2\rightarrow 2}\times 12g^2\frac{1}{k^2_\perp}\left(1+\frac{M_{HQ}^2}{s}e^{2\eta}\right)^{-2}
\end{align}

{  $g$ is a running coupling constant; for more detail, we refer to Ref.\cite{PhysRevD.71.114510} as,

\begin{align*}
    g^{-2} = 2\beta_0 \ln \left(\frac{\mu T}{\Lambda_{\text{MS}}} \right) + \frac{\beta_1}{\beta_0} \ln \left(2 \ln \left(\frac{\mu T}{\Lambda_{\text{MS}}} \right)\right),
\end{align*}
with
\begin{align*}
\beta_0 = \frac{1}{16\pi^2}\left(11 - \frac{2}{3} N_f\right)
\end{align*}

\begin{align*}
\beta_1 = \frac{1}{(16\pi^2)^2}\left(102 - \frac{38}{3} N_f\right)
\end{align*}

where $\mu = 2\pi$, $\Lambda_{\text{MS}} = T_c/0.77$, $T_c$ = 0.155 GeV}. $\eta$ is the rapidity of emitted gluons and $\big(1+\frac{M_{HQ}^2}{s}e^{2\eta}\big)^{-2}$ is the dead cone factor\cite{ALICE:2021aqk}. As HQs traverse the medium, the dead-cone effect may suppress their energy loss \cite{Dokshitzer:2001zm}. Next, the energy and transverse momentum of the emitted gluon can be described in terms of the rapidity variable as,

\begin{align}
\label{31}
    &E_5=k_\perp\cosh (\eta), &&k_z=k_\perp\sinh (\eta),
\end{align} 
and 
\ba 
d^3{ k}_5=d^2{ k}_\perp dk_z=2\pi k_\perp^2 dk_\perp \cosh(\eta) d\eta.
\ea

The connection between the interaction time and the interaction rate ($\Lambda$) is taken by the limitations on the $\tau_F$, \cite{Das:2010tj} as $\tau=\Lambda^{-1}>\tau_F=\frac{\cosh(\eta)}{k_\perp}$. This can be precisely written as, $k_\perp>\Lambda\cosh(\eta)=(k_\perp)_{\text{min}}$, where $(k_\perp)_{{min}}$ is the minimum value of $k_{\perp}$. The second constraint on the  $E_5$ is given as $\text{E}>E_5=k_\perp\cosh(\eta)$. Finally, one can write the transport coefficients for the radiational processes in terms of the collisional transport coefficient, $X_{c}$ of the HQs given in Eq.~\eqref{24} as,
{\small 
\begin{align}
\label{trans_rad}
    X_r = X_c \frac{6\alpha}{\pi}\int\int_{(k_\perp)_{min}}^{(k_\perp)_{max}} \frac{dk_\perp d\eta}{k_\perp} \frac{T}{k_\perp \cosh \eta} \left( 1+\frac{M_{HQ}^2}{s}e^{2\eta}\right)^{-2},
\end{align}}
where $(k_\perp)_{{max}}=\frac{\text{E}}{\cosh(\eta)}$. Next, we shall discuss the contribution through medium polarization. 

\subsection{Energy Loss through Medium Polarization}
\label{sec:HQP}

In the domain of polarization energy loss, we assume that the energy of the moving HQ is significantly higher than the energy loss incurred during its propagation. In this case, the energy loss of HQs per unit length can be obtained from the Wong equations~\cite{Wong:1970fu} and given as \cite{10.1088/1361-6471/ad10c9},

            \begin{equation}
                    \frac{{\text {dE}}}{{\text {dx}}}\bigg|_\mathrm{pol}=g~q^{a}~ \frac{{v^i}}{{|\bf v|}}
                    { \it E}^{a,i}_{ind}({\bf x},t)
                    \label{pol1}
            \end{equation}
where $g$ is strong coupling, $q^a$ is color charge, ${\text E}$ is the energy and { $v=\frac{{p}}{\sqrt{{p}^2+M_{HQ}^2}} $ is the velocity }of the HQs. ${ \it E}^{a,i}_{ind}({\bf x},t)$ is the induced color-electric field resulting from the HQ motion that can be obtained by solving the Yang-Mills equations for a thermalized gluonic system with the source given by the color current carried by the HQ in the Fourier space as ~\cite{YousufJamal:2019pen},

\ba
{E}^{a,i}_{ind}(K)=i\omega \Delta^{ij}(K)j^{a,j}_{ext}(K),
\label{eq:eind}
\ea
where, $K \equiv k_\mu=({\omega,{\bf k}})$ and $\Delta_{ij}(K)$ is the gluon propagator. The external current is given as,
\ba
j^{a,j}_{ext}(K)=\frac{i g q^a v^j}{\omega -{ k}\cdot{ v}+i 0^+}.
\label{eq:jind}
\ea
The gluon propagator can be obtained as ~\cite{Jamal:2017dqs, Kumar:2017bja},

\ba
\Delta_{ij}(K)= [k^2\delta_{ij}-k_{i}k_{j} -\omega^2\epsilon_{ij}(K)]^{-1},
\label{eq:pro}
\ea
where $\epsilon_{ij}(K)$ is the medium dielectric permittivity that can be expanded in the transverse $A^{\text{ij}}$ and longitudinal $B^{\text{ij}}$ projections as,
\ba
    \epsilon^{ij}(K)= A^{ij}~\epsilon_T(K)+B^{ij}~\epsilon_L(K),
    \label{eq:eplt1}
\ea  
where,  $A^{\text{ij}}=\delta ^{\text{ij}}-\frac{k^i k^j}{k^2}$, and $ B^{\text{ij}}=\frac{k^i k^j}{k^2}$.            
Next, using Eq.\eqref{eq:eind} with Eq.\eqref{eq:jind}, \eqref{eq:pro} and \eqref{eq:eplt1}, the energy loss due to medium polarization given in Eq.~\eqref{pol1} can be written as,

\begin{align}
\frac{{\text {dE}}}{{\text {dx}}}\bigg|_\mathrm{pol} &= -\frac{C_F \alpha_s}{2 \pi ^2 |{ v}|^2}
\int^{k_{\text{max}}}_{k_0} d^3k\frac{\omega }{k^2}\bigg\{\left(k^2 |{ v}|^2-\omega ^2\right)\notag\\
&\times\text{Im}\Big[\frac{1}{\omega ^2 \epsilon_T(K)-k^2}\Big]+\text{Im}\Big[\frac{1}{\epsilon_{L}(K)}\Big]\bigg\}_{\omega = k \cdot  v}.
\label{eq:de}
\end{align}
Here, $C_F =4/3$ is the Casimir invariant in the fundamental representation of $SU(3)$ and { $\alpha_s=g^2/4\pi$ represents the QCD running coupling constant at finite temperature. Equation~\eqref{eq:de} describes the energy change of the HQs due to medium polarization, where the effects of the medium and temperature are encoded in $\epsilon_{L}(\omega,{\bf k})$ and $\epsilon_{T}(\omega,{\bf k})$. Here, we provide a brief derivation of $\epsilon_{L}(\omega,{\bf k})$ and $\epsilon_{T}(\omega,{\bf k})$ with the necessary steps. For detailed derivations and further insights, we recommend consulting Refs.~\cite{Jamal:2020fxo, Jamal:2021btg, Jamal:2020emj}.
For the background bulk/ QGP medium, we begin with solving the Boltzmann transport equation given as,
\begin{align}
\label{eq:BM}
u^{\mu }\partial_{\mu}^{x}\delta f_a^i(p,x) + g\theta_i u_{\mu} F^{\mu \nu}_a(x) \partial_{\nu}^p f^{i}_{0}(p) = 0,
\end{align}

where 

\begin{align}
f^{i}_a(p,X) = f^{i}_{0}(p) + \delta f^{i}_a(p, X),
\end{align}

with $f^{i}_{0}(p)$ being the isotropic distribution function of medium particles, and $\delta f^{i}_a(p, X)$ representing the small perturbation due to the motion of HQs such that

\begin{align}
f^{i}_{0}(p) \gg \delta f^{i}_a(p, X),
\end{align}

where ${\text u}^{\mu}$ is the velocity four-vector, $F_{a}^{\mu\nu}$ is the chromodynamic field strength tensor, and $\theta _{i}$ takes the value $1$ for quarks and gluons, and $-1$ for anti-quarks. Upon determining $\delta f^{i}_a(p, X)$, we compute the induced current given by

\ba
\label{eq:jind}
j_{ind,a}^{\mu}(X) &=& g \int \frac{d^{3}p}{(2\pi)^3} u^{\mu} \Big\{ 2N_c \delta f^{g}_a(p,X) + N_{f}[\delta f^{q}_a(p,X) \nn 
&-& \delta f^{\bar{q}}_a(p,X)] \Big\},
\ea

where $N_{c}$ is the number of color charges and $N_{f}$ is the number of quark flavors. The polarization tensor, $\Pi^{ij}(K)$, is extracted from the induced current in Fourier space as

\begin{align}
\Pi^{ij}(K) = \frac{\delta J^{i}_{a, ind}(K)}{\delta A_{j,a}(K)}.
\label{eq:perm}
\end{align}
The $\epsilon^{ij}(K)$ tensor of the hot QGP medium is obtained from $\Pi^{ij}(K)$ as

\begin{align}
\epsilon^{ij}(K) = \delta^{ij} - \frac{1}{\omega^2} \Pi^{ij}(K).
\label{eq:ep}
\end{align}
By solving Eq.~\eqref{eq:ep} with proper tensor contraction, we derive $\epsilon _L(K)$ and $\epsilon _T(K)$ as follows:

\begin{align}
\epsilon _L(K) &= 1 + m_D^2 \left[ \frac{1}{k^2} - \frac{\omega}{2k^3} \ln \left( \frac{\omega+k }{\omega-k} \right) \right],
\label{eq:eL}
\end{align}
and
\begin{align}
\epsilon _T(K) &= 1 - \frac{m_D^2}{2 \omega k} \left[ \frac{\omega}{k} + \left( \frac{1}{2} - \frac{\omega^2}{2k^2} \right) \ln \left( \frac{\omega+k }{\omega-k} \right) \right].
\label{eq:eT}
\end{align}
Here, the screening mass, $m_D$, is defined in terms of the equilibrium distribution functions (Bose-Einstein, $f_{g} $ and Fermi-Dirac, $f_{q,\bar{q}} $) as
\ba
\label{eq:debye1}
m_D^2 &=& -4 \pi \alpha_s \bigg( 2 N_c \int \frac{d^3 p}{(2 \pi)^3} \partial_p f_g ({\bf p}) \nn
&+& N_f \int \frac{d^3 p}{(2 \pi)^3} \partial_p (f_q({\bf p}) + f_{\bar{q}}({\bf p})) \bigg).
\ea
where the distribution functions are given as, 
	\ba
	\label{eq1}
	f_{g}= \frac{\exp[-\beta E_g]}{1- \exp[-\beta E_g]},~~~~
	f_{q,{\bar q}}= \frac{\exp[-\beta E_q]}{1+ \exp[-\beta E_q]}.
	\ea
Here, $\beta = 1/T$, the energy of gluons, $E_{g}=|{\bf q}|$  and for quarks/ antiquarks $E_{q} = \sqrt{|{\bf q}|^2+m_q^2}$ (${\bf q}$ and $m_q$ denotes their momenta and masses of the medium particles, respectively). In the limit $T\gg m_q$, we neglect the mass of the light quarks.}

This process of polarization energy loss helps in understanding the complex interplay between the HQ and the surrounding medium.  Additionally, the contribution of elastic and inelastic collisions can be obtained using Eq.\eqref{24} and Eq.\eqref{trans_rad}, respectively. These three processes can cause the HQs to lose energy in the QGP medium. In the following section, we will discuss the energy loss and drag coefficients for each of these processes, along with their potential influence on the $R_{AA}$ of the HQs.

\begin{figure*}
 \centering
\includegraphics[scale=0.32]{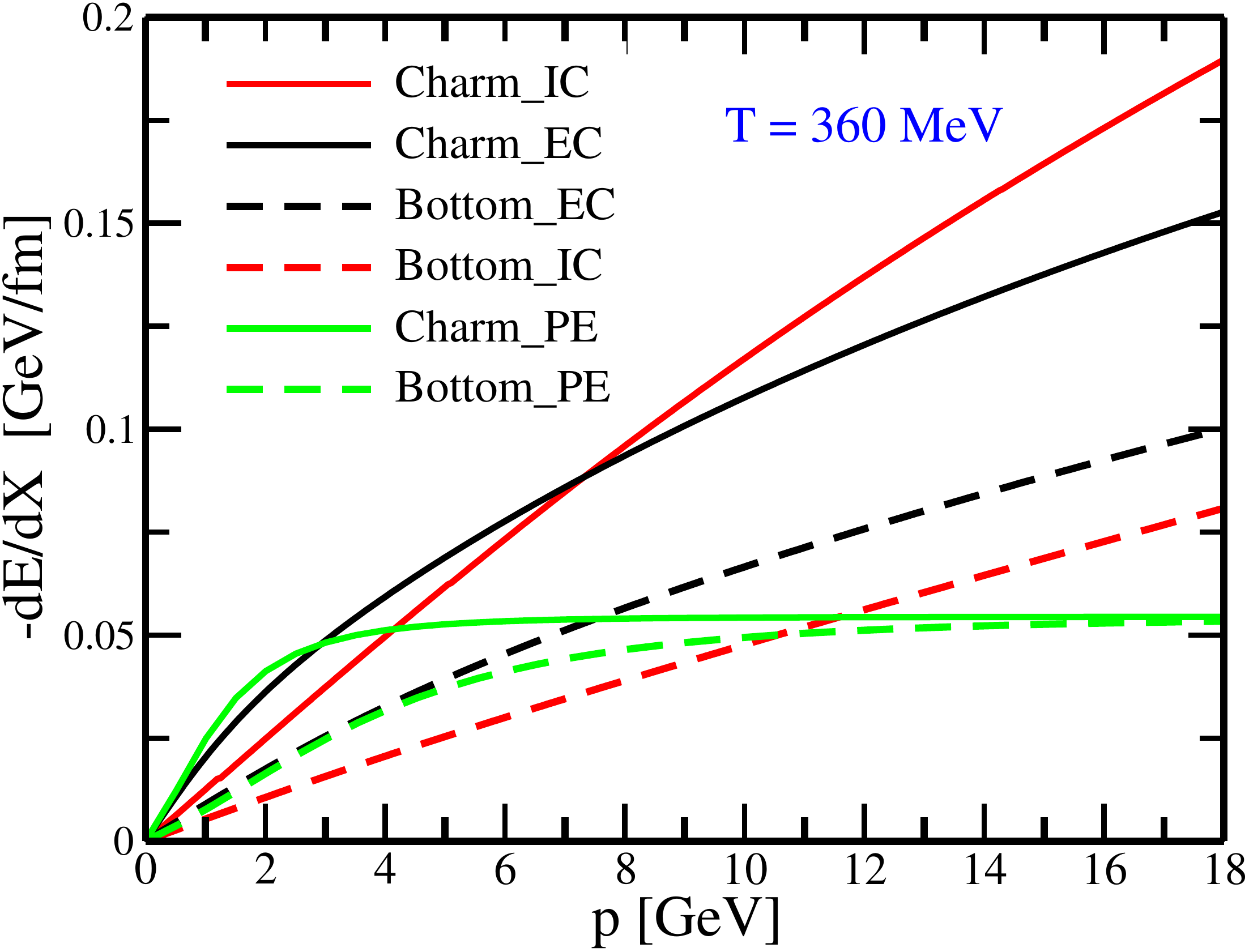}
 \hspace{1 cm}
 \includegraphics[scale=0.32]{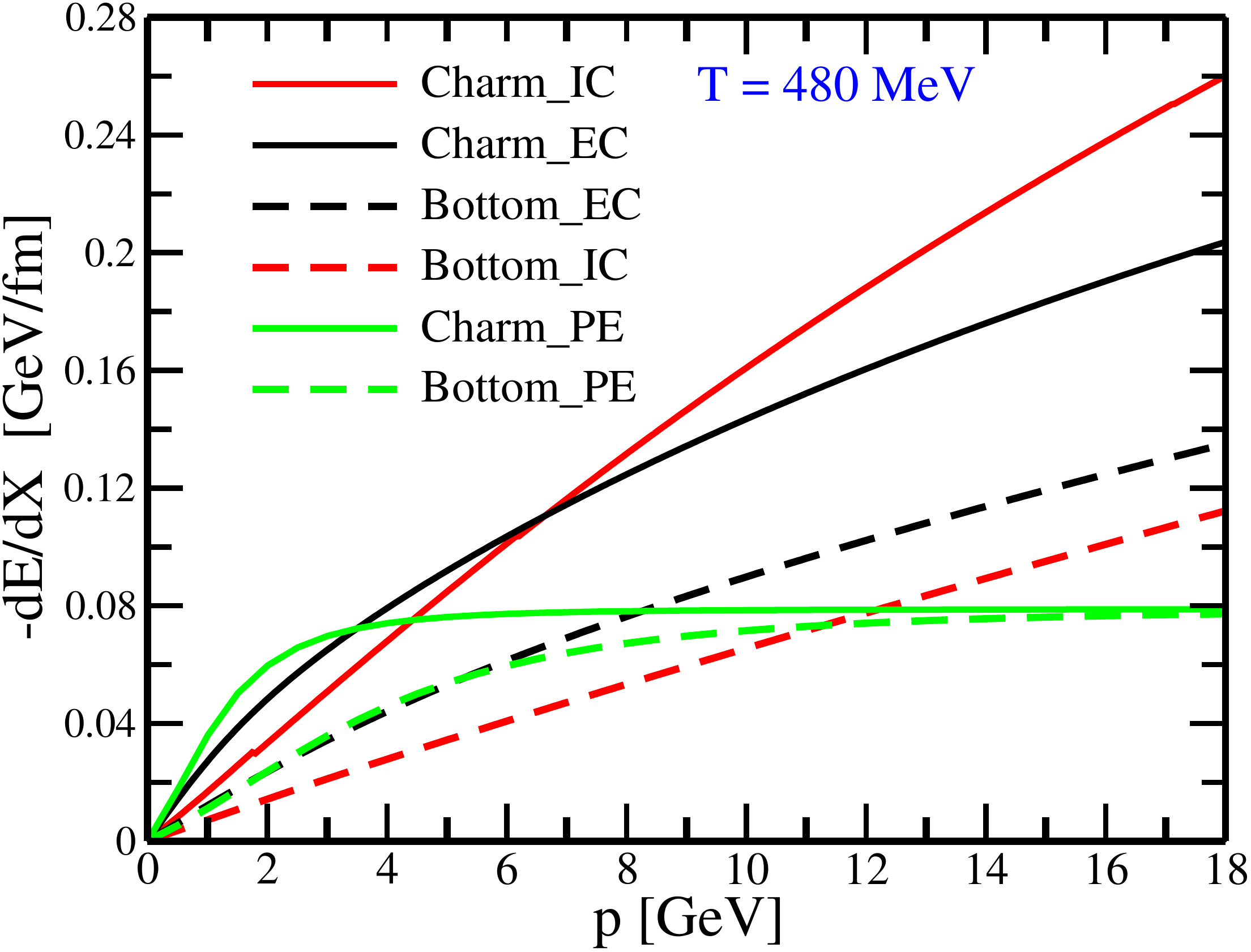}
\caption{ Comparing the momentum dependence of the HQs energy loss from polarization effect (PE), elastic collisions (EC), and radiation (IC) for the charm and bottom quark for T = 360 MeV (left panel) and  T = 480 MeV (right panel).}
\label{E_p}
\end{figure*}

\begin{figure*}
 \centering
\includegraphics[scale=0.32]{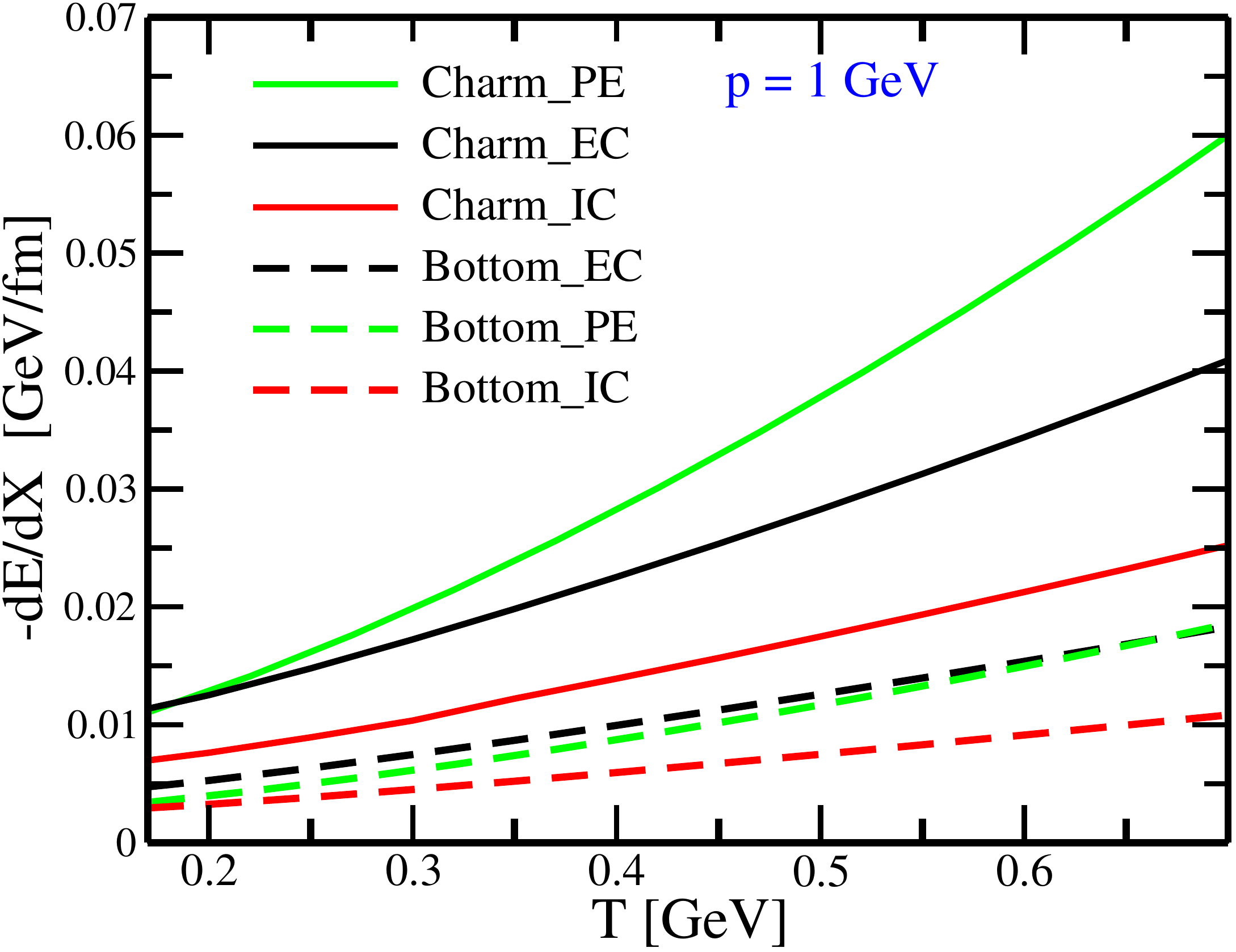}
\hspace{1cm}
\includegraphics[scale=0.32]{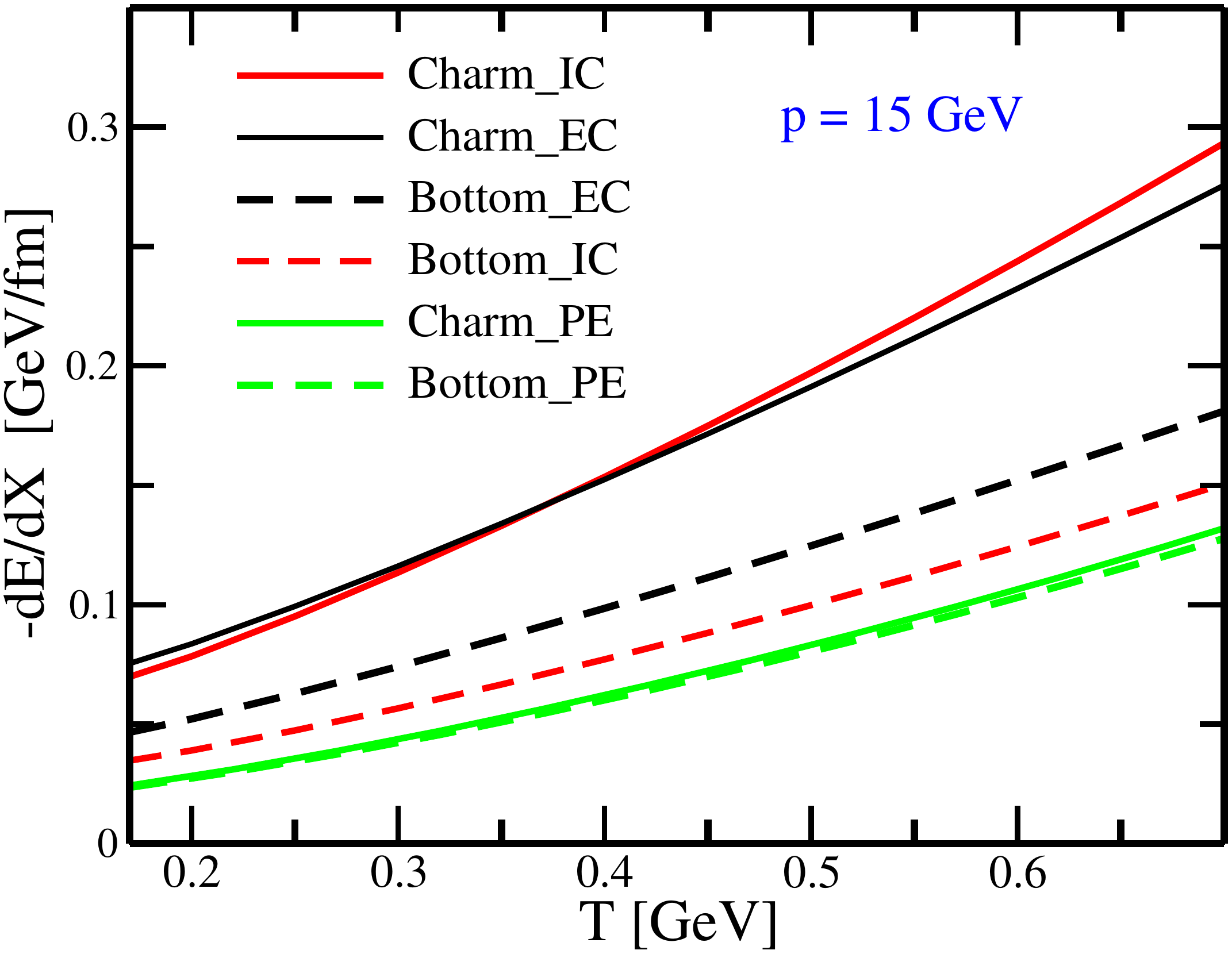}
\caption{ Comparing temperature dependence of the HQs energy loss from PE, EC, and IC for the charm and bottom quark at p = 1 GeV (left panel) and p = 15 GeV (right panel).}
\label{E_t}
\end{figure*}

\begin{figure*}
 \centering
\includegraphics[scale=0.32]{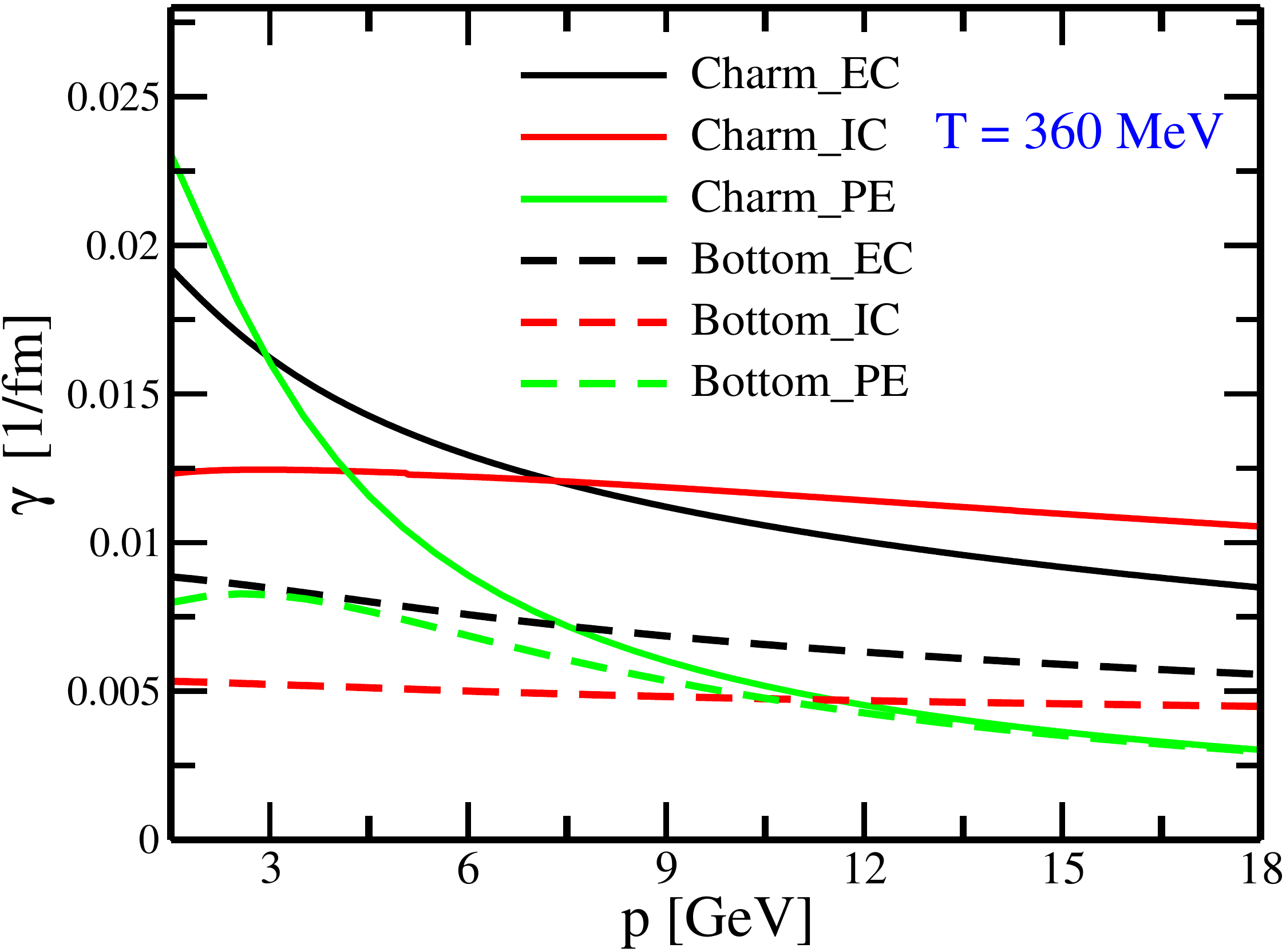}
 \hspace{1 cm}
 \includegraphics[scale=0.32]{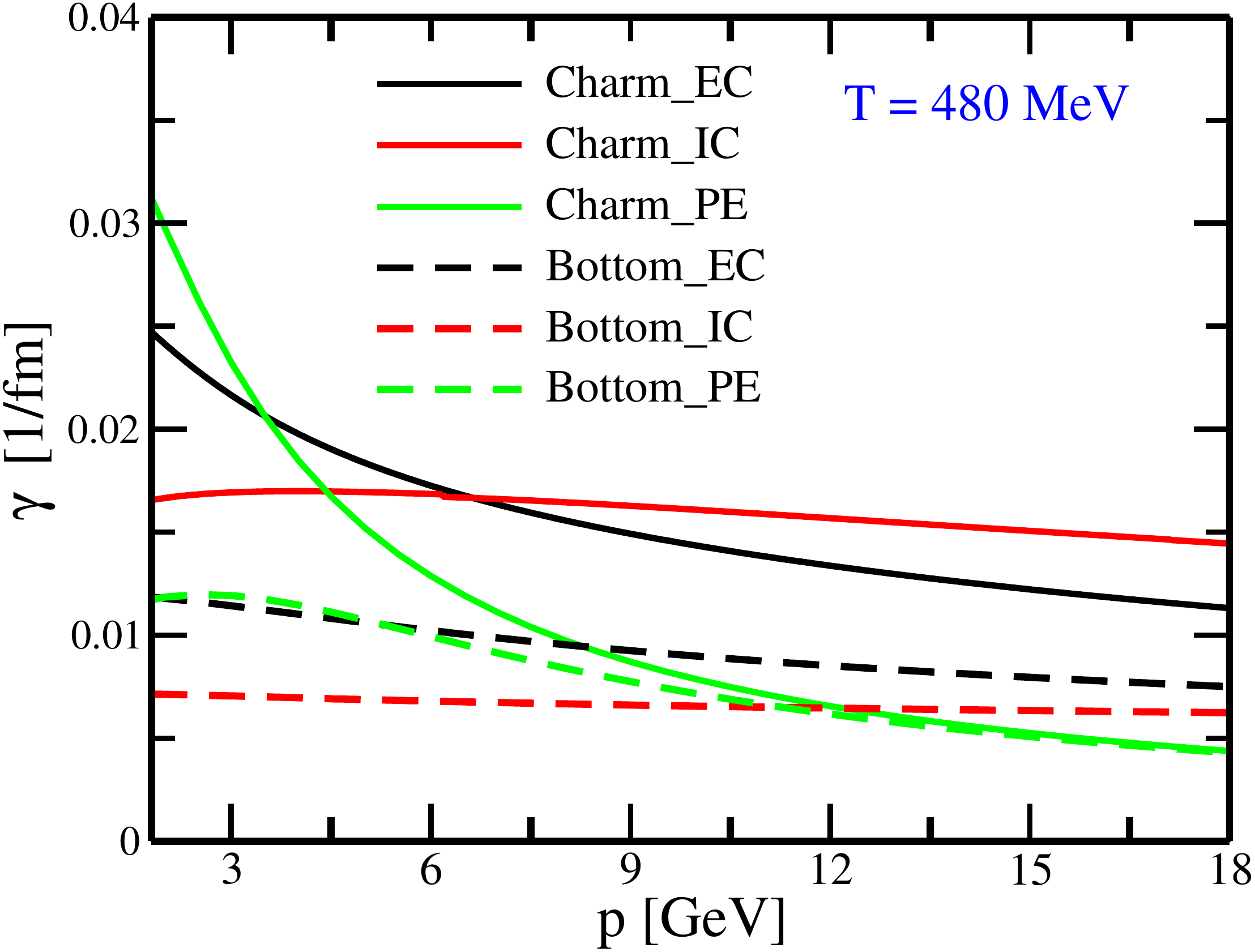}
\caption{ Comparing the momentum dependence of the HQs drag coefficient from PE, EC, and IC for the charm and bottom quark for T = 360 MeV (left panel) and  T = 480 MeV (right panel).}
\label{drag_p}
\end{figure*}

\begin{figure*}
 \centering
\includegraphics[scale=0.32]{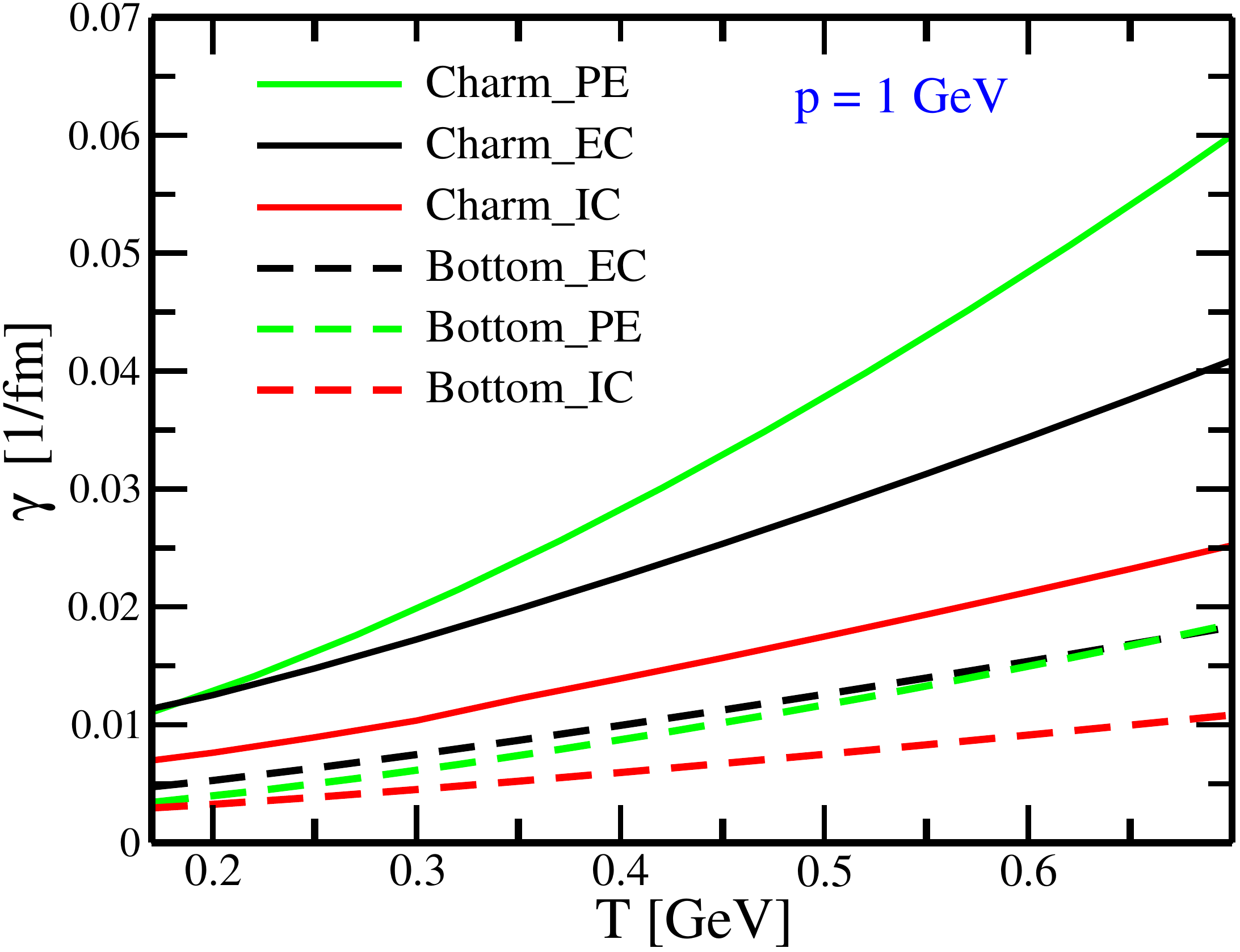}
\hspace{1 cm}
\includegraphics[scale=0.32]{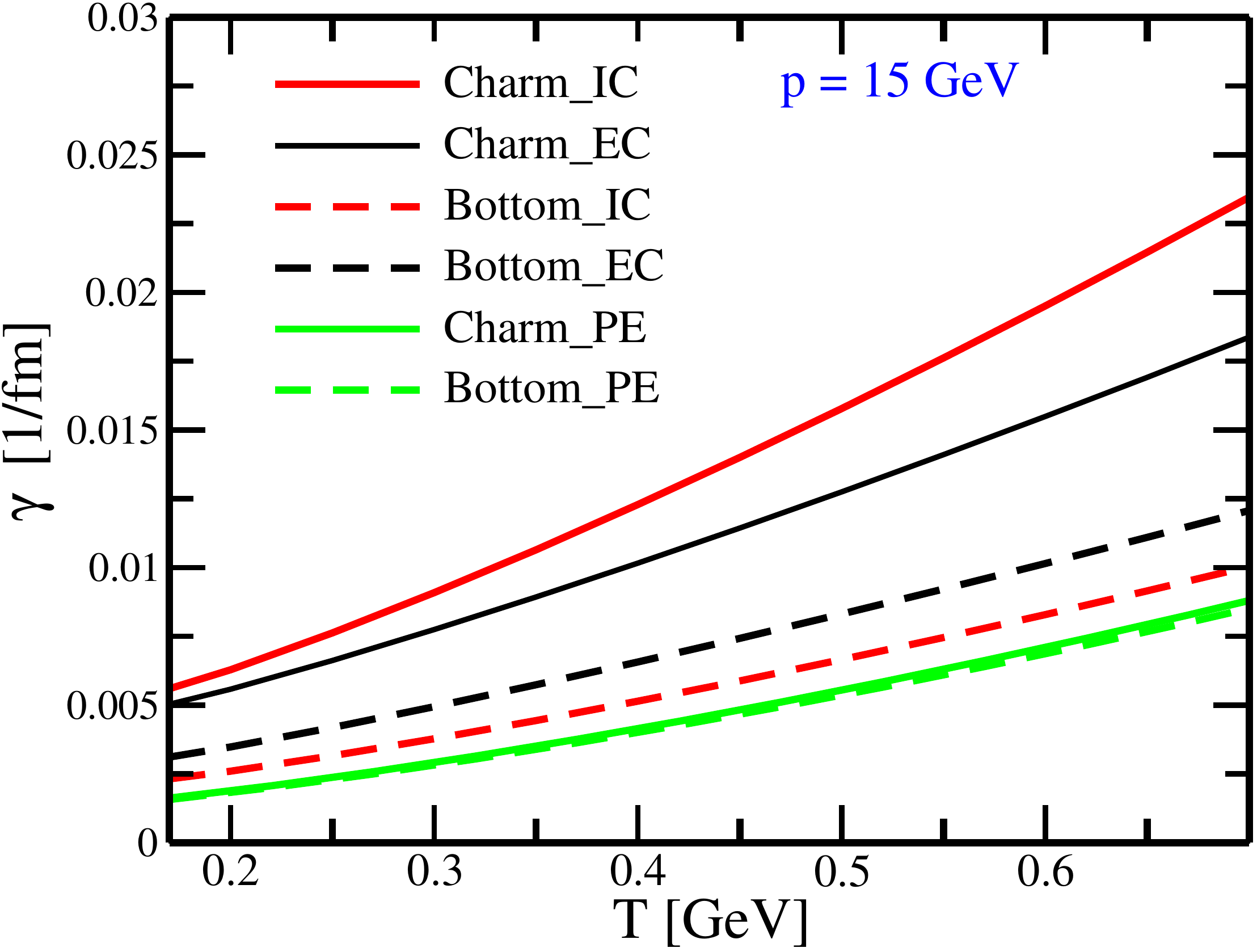}
\caption{ Comparing the temperature dependence of the HQs drag coefficient from PE, EC, and IC for the charm and bottom quark at p = 1 GeV (left panel) and p = 15 GeV (right panel).}
\label{drag_t}
\end{figure*}

\section{Results and Discussions}
\label{RD}

To carry out a comparative study, we have plotted three quantities separately from all three processes, namely the energy loss ($-{\text {dE}}/{\text {dx}}$), the drag coefficient ($\gamma$), and the experimental observable, the $R_{AA}$. The results from the medium polarization case are denoted as ``PE" (represented by the green color), whereas ``EC" (represented by the black color) indicates the elastic collision. The inelastic collisions (represented by the red color) are denoted as ``IC". The results of charm quarks are represented by solid curves, whereas the results of bottom quarks are represented by dashed curves. The masses of charm quarks ($M_c$) and bottom quarks ($M_b$) are considered to be $1.5$ GeV and $4.5$ GeV, respectively. The comparison is based on two important parameters, namely, the HQ momentum (${p}$) and medium temperature ($T$).  To illustrate the variations with the HQ momentum, we are showing the results at T=0.360 GeV and T=0.480 GeV, corresponding to the initial temperature of the center of the fireball at RHIC and LHC energies, respectively. On the other hand, when demonstrating the variation of the results with respect to the medium temperature, the values of the HQ momentum remain fixed at $p=1$ GeV and $p=15$ GeV. It should be noted that the results for PE and IC have already been discussed by some of us in Ref. \cite{10.1088/1361-6471/ad10c9}. Therefore, this discussion will primarily focus on the EC results and subsequently compare them with the other two cases.

\subsection{Energy Loss}

The plots in Fig.~\ref{E_p} illustrate the variation of energy loss of the HQs with respect to  $p$, for all three processes—energy loss due to polarization, elastic collisions, and inelastic collisions—at fixed temperatures using Eqs.~\eqref{eq:de}, ~\eqref{24}, and ~\eqref{trans_rad}, respectively. It is worth noting that the energy loss through all three processes increases as the momentum increases. However, charm and bottom quarks do not follow the same trend in all three cases.
At T = 360 MeV (left panel of Fig.~\ref{E_p}), the energy loss of the charm quark due to polarization dominates at low momentum. However, in the case of the bottom quark, elastic collisions dominate throughout, followed by polarization at low momentum and radiation loss. Moreover, the polarization loss saturates at higher momentum, as the velocity of the HQ, $v \rightarrow 1$, and the process becomes independent of $p$. On the other hand, elastic collision and radiation increase with momentum.
The charm quark's radiation surpasses elastic collision at sufficiently high momentum, $p \sim$ 7 GeV, giving the highest contribution. This phenomenon highlights the fascinating interplay of energy loss mechanisms at extreme momentum regimes. The HQ mass has a substantial impact on the ``dead cone" effect, as described by the factor $\left(1+\frac{M_{HQ}^2}{s}e^{2\eta}\right)^{-2}$ given in Eq.~\eqref{rad}. This effect becomes more pronounced with increasing HQ mass, resulting in reduced radiation loss for heavy bottom quarks compared to lighter charm quarks, also documented in recent research work \cite{ALICE:2021aqk}.
At such a high momentum, the charm quark dead cone effect gets suppressed. However, the bottom quark radiation loss still lacks energy loss due to the elastic scattering process, even at $p\sim$ 18 GeV. Hence, in this case, elastic collision gives the dominant contribution.
In the right panel of Fig.~\ref{E_p}, the results are shown for a higher temperature, T = 480 MeV. We observed a similar pattern in all three processes, but the magnitude of loss has increased in each case.

To better understand the charm and bottom energy loss dependence on temperature, we show the energy loss versus temperature for all three processes at two different values of momentum, $p$ = 1 GeV (left panel) and 15 GeV (right panel) in Fig.~\ref{E_t}. We have chosen these values because the results mostly change as we move within this range of momentum, as noticed in Fig.~\ref{E_p}. It is observed that the energy loss in all three cases increases with both momentum and temperature.
In the left panel of Fig.~\ref{E_t}, the polarization loss of the charm quark dominates for the full range of temperatures considered, followed by the elastic and then inelastic loss. However, this is not the case for the bottom quark, where elastic collision dominates overall. But at higher momentum, $p$ = 15 GeV, as shown in the right panel of Fig.~\ref{E_t}, the energy loss due to polarization gives the least contribution for both charm and bottom quarks. The overlap of charm quark and bottom quark energy loss at high momentum due to medium polarization arises because both get saturated to nearly the same value at such high momenta, $p\gg M_Q$. Again, there is an interplay between the results of energy loss due to elastic and inelastic scattering for charm quark. Initially, the energy loss due to elastic scattering dominates. Nonetheless, at $p$ = 15 GeV and T $\sim$ 0.35 GeV, the radiation loss starts dominating. Nevertheless, this is different for bottom quarks, where the elastic scattering contribution dominates throughout, even at $p$ = 15 GeV.

\subsection{Drag coefficient}

Exploring the dynamics of energy loss from an alternative standpoint, we have conducted a comprehensive analysis of the drag experienced by the HQs in the QGP medium. The relationship between energy loss and drag is expressed through the equation $\gamma=-\frac{1}{p}\left(\frac{{\text {dE}}}{\text {dx}}\right)$ \cite{Debnath:2023zet, Prakash:2023zeu}, signifying that the energy loss is suppressed by the HQ momentum. In Fig.~\ref{drag_p}, we present the drag coefficient plotted against momentum at a constant temperature for three distinct processes, revealing a notable decrease with increasing momentum. Crucially, a diminished drag coefficient signifies a weakened interaction between the HQ and the surrounding medium.
The investigation highlights intriguing variations in the drag coefficient among different processes, with the polarization process exhibiting a rapid change compared to elastic collision and radiation. As depicted in Fig.~\ref{E_p}, the energy loss due to radiation follows almost linear trend with momentum, while the presence of the $\frac{1}{p}$ factor in $\gamma$ results in a nearly constant or slowly changing behavior. The distinct behavior of $\gamma$ for polarization is evident as it steeply decreases initially and then slows down, avoiding saturation due to the $1/p$ factor. In Fig.~\ref{drag_p} (left panel), the drag coefficients exhibit comparable magnitudes across the three cases, transitioning at different momentum regimes. For instance, in the case of charm quarks, the polarization case starts with the highest magnitude, gradually reaching the lowest values as momentum increases. A similar trend is observed for bottom quarks, where the coefficient due to elastic collision dominates throughout. In Fig.~\ref{drag_p} (right panel), at a higher temperature of T = 480 MeV, the overall pattern in all three processes remains similar, albeit with an increased magnitude of drag at the same momentum compared to T = 360 MeV.

To gain a broad understanding of $\gamma$'s temperature dependence, we have plotted $\gamma$ against T at fixed momenta of $p$ = 1 GeV and $p$ = 15 GeV, as shown in the left and right panel of Fig.~\ref{drag_t}, respectively. The results indicate an increase in the drag coefficient with temperature at both fixed moments. At low momentum in the left panel, there is a dominance of polarization in the drag coefficient for charm quarks, followed by elastic collision, then radiation. Conversely, for bottom quarks, polarization, and elastic scattering effects overlap, overshadowing radiation. In the right panel, at $p$ = 15 GeV, the dominance of radiation for charm quarks is evident, whereas for bottom quarks, elastic collision effects take precedence, followed by radiation and then polarization. The overlap in the polarization of charm and bottom quarks at high momentum is attributed to their saturation at nearly identical values. It is noteworthy that the magnitude of $\gamma$ at higher momentum is substantially reduced.

\begin{figure}
 \centering
\includegraphics[scale=0.32]{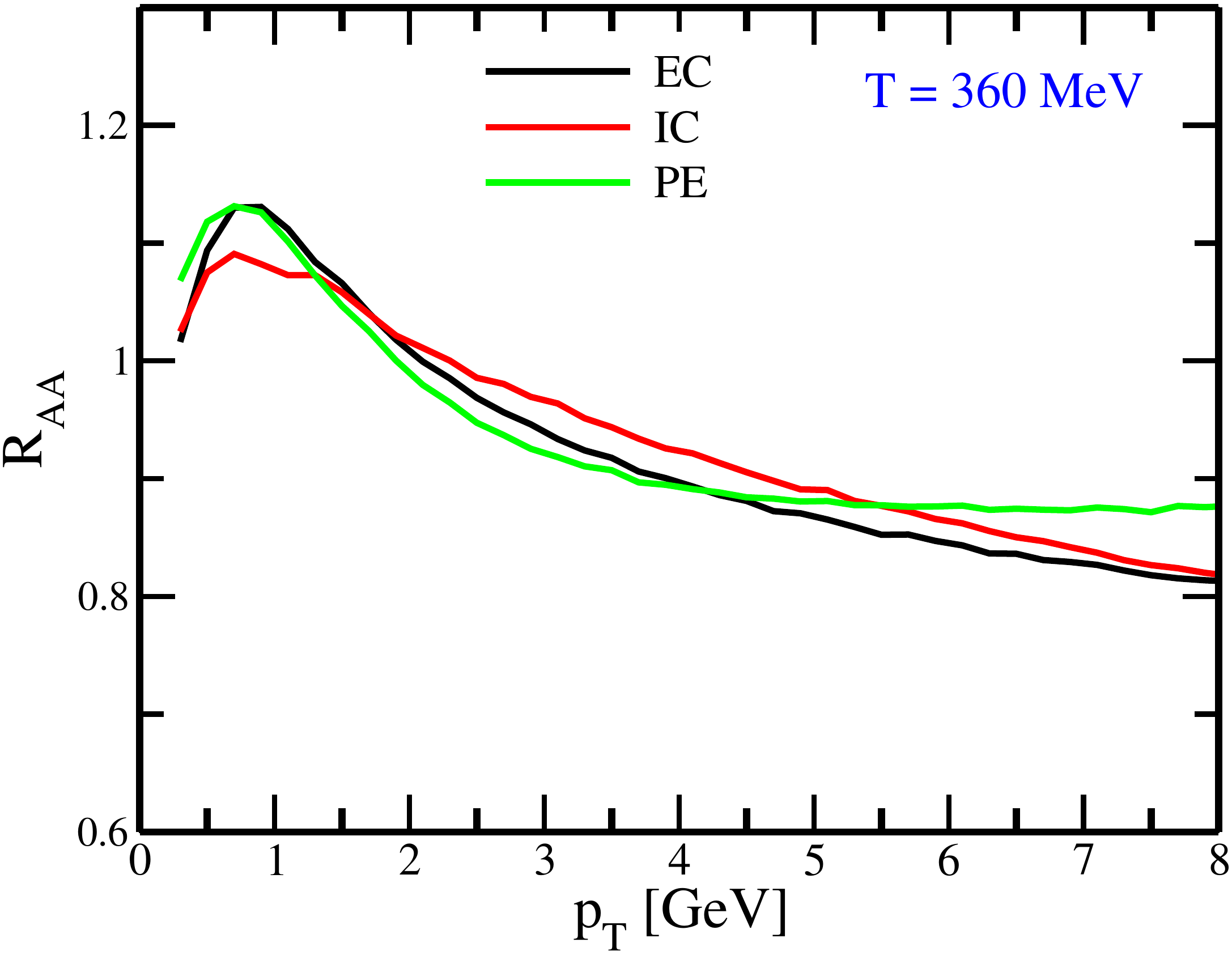}
 \hspace{1 cm}
 \includegraphics[scale=0.32]{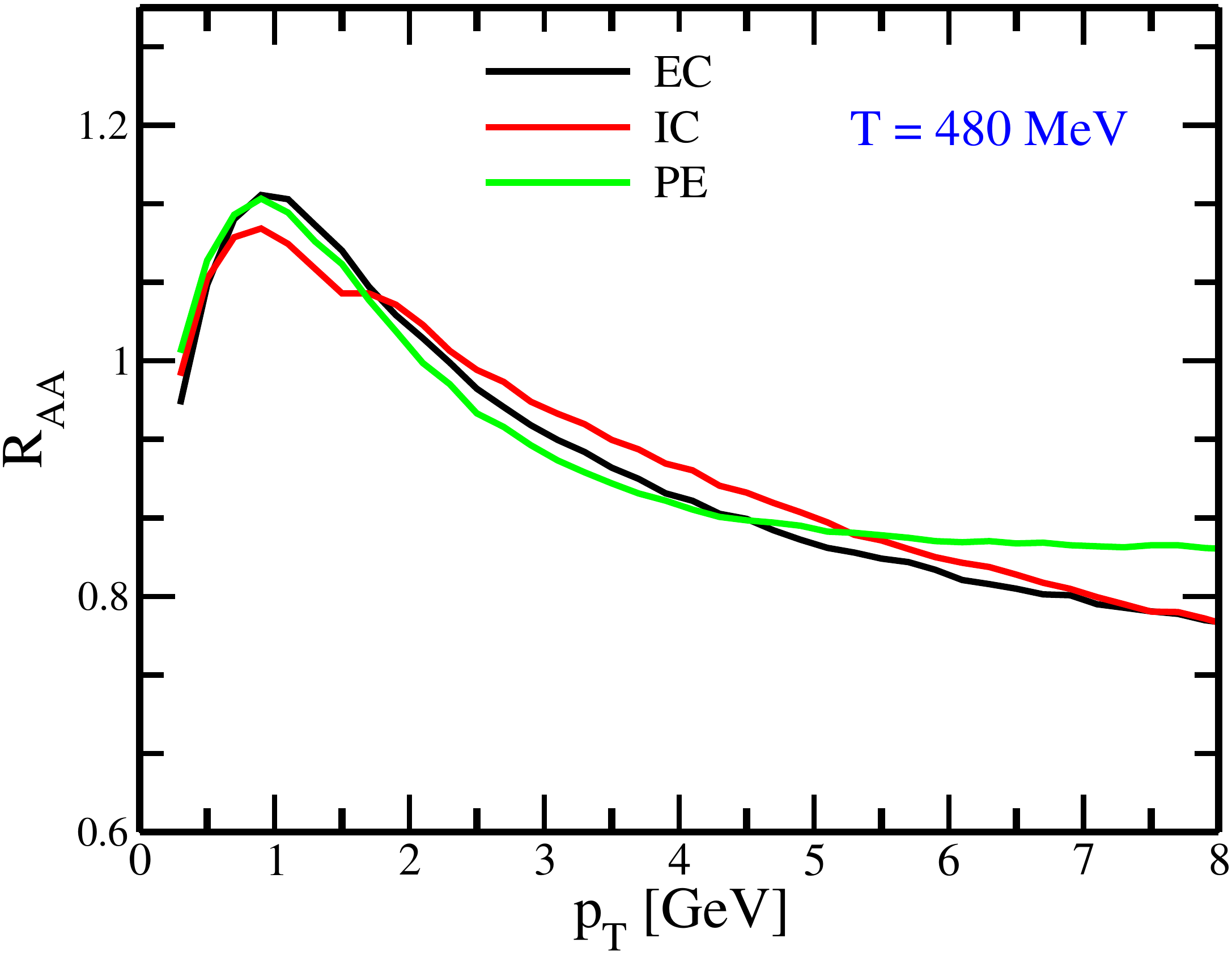}
\caption{ The $R_{AA}$ of the charm quark as a function of  $p_T$ for to PE, EC, and IC at T = 360 MeV (upper panel)  and at T = 480 MeV (lower panel).}
\label{raa}
\end{figure}

\begin{figure}
 \centering
\includegraphics[scale=0.32]{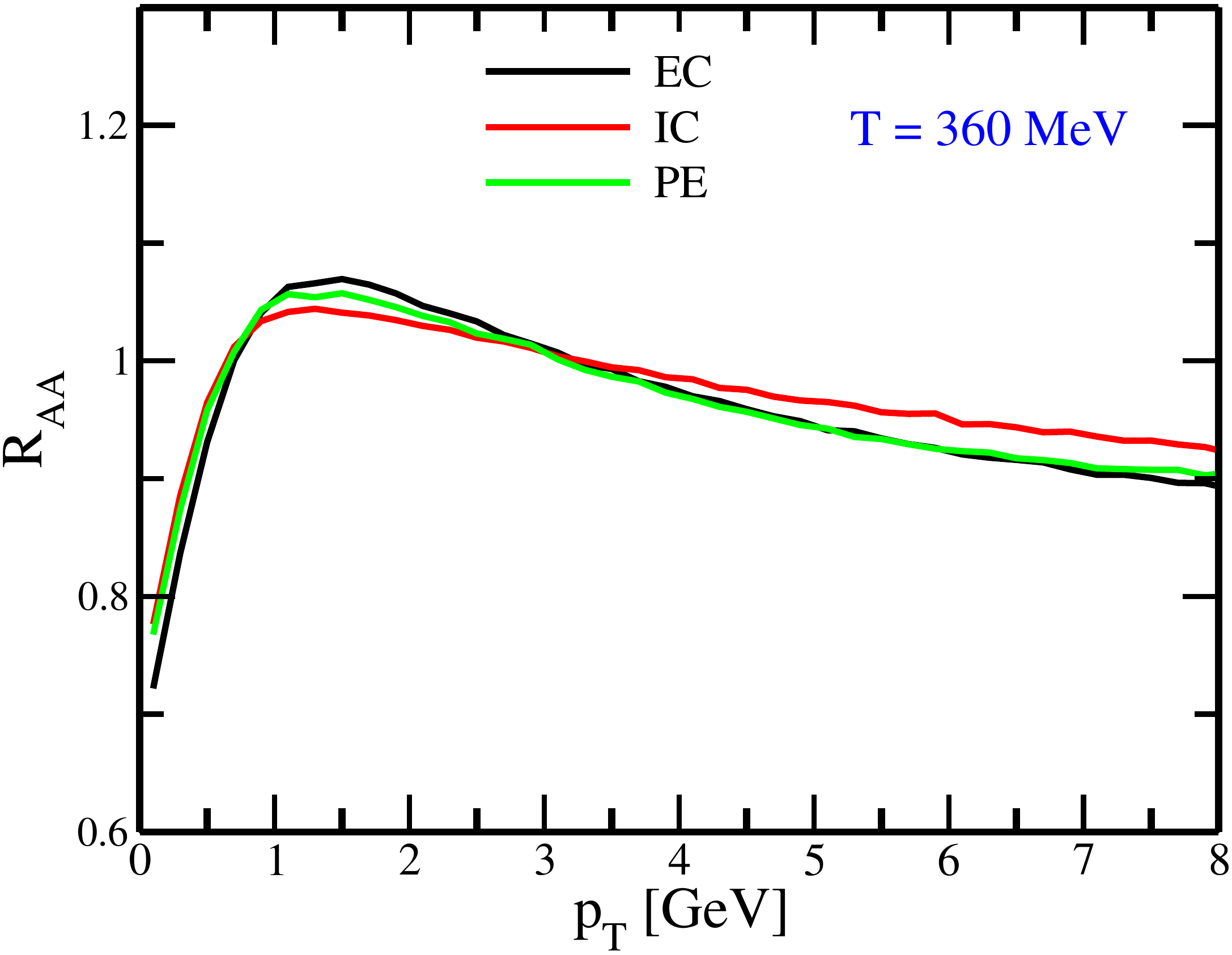}
 \hspace{1 cm}
 \includegraphics[scale=0.32]{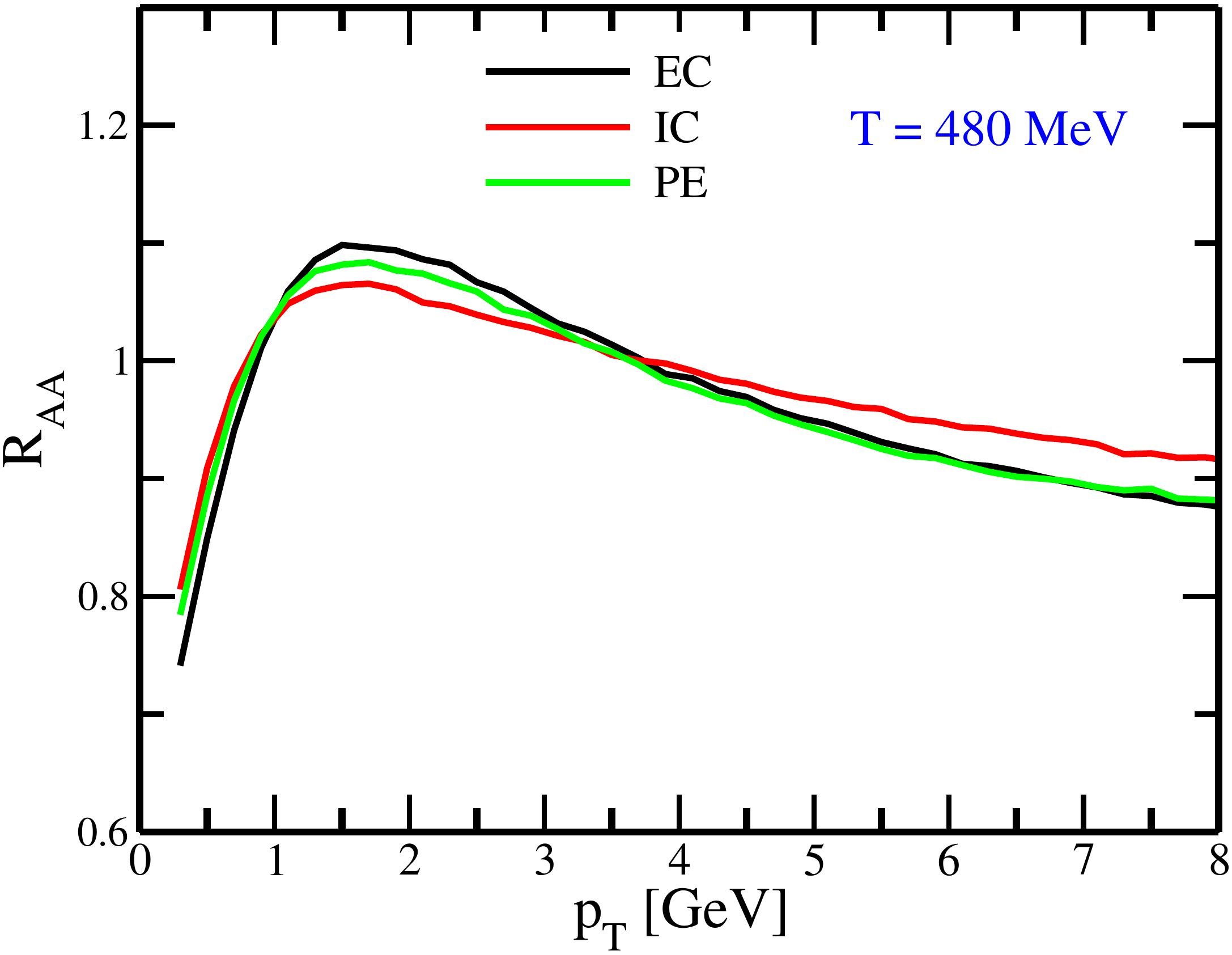}
\caption{ The $R_{AA}$ of the bottom quark as a function of  $p_T$ for to PE, EC, and IC at T = 360 MeV (upper panel)  and at T = 480 MeV (lower panel).}
\label{raa_b}
\end{figure}

\subsection{Nuclear modification factor, $R_{AA}$}
\label{sec:raa}
The  $R_{AA}$ is an important measurement in HIC experiments. It is defined as the ratio between the initial momentum spectra, $f_{\tau_0} (p_T)$, and the final momentum spectra, $f_{\tau_f} (p_T)$, at time ${\tau_f}$. When $R_{AA}$ approaches unity, it means that the HQs have negligible interaction with the surrounding bulk medium. The distribution function $f_{\tau_f}(p)$ is determined at an evolution time of $\tau_f = 7$ fm/c, while $f_{\tau_0}$ is estimated within the framework of fixed order + next-to-leading log (FONLL) QCD corrections ~\cite{Cacciari:2005rk, Cacciari:2012ny}. This approach has been successful in reproducing the spectra of D-mesons in $p-p$ collisions following the fragmentation process, which can be parameterized as: 
\ba \frac{dN}{d^2p{_T}} = \frac{x_0}{(x_1+p_T)^{x_2}}
\ea
where $x_0= 6.365480 \times{10}^8$, $x_1= 9.0$ and $x_2= 10.27890$.
To calculate the $R_{AA}(p_T)$ of the HQs in a hot QCD medium, the Langevin dynamics is used and the corresponding equations are given as follows,
\begin{align}
&dx_i=\frac{p_i}{\text E} dt,\nonumber \\
&dp_i=-\gamma p_i\, dt+C_{ij}\rho_j\sqrt{dt},
\end{align}
where $dt$ is the time step and $dp_i$ and $dx_i$ represent the change in momentum and position of the HQs, respectively. The HQs undergo random motion within the medium, influenced by two distinct forces: the dissipative force, which contains the  $\gamma$, and the stochastic force, which governs the thermal noise ($\rho_j$). The covariance matrix $C_{ij}$ can be expressed as $C_{ij} = \sqrt{2D}\delta_{ij}$ in the static limit $p \rightarrow 0$. This diffusion coefficient is estimated from drag coefficients using Einstein's relation, given as $D = \gamma {\text E} {\text T}$ \cite{Walton:1999dy, Mazumder:2013oaa, Moore:2004tg}.

In Fig.~\ref{raa}, we present the plots of $R_{AA}$ for charm quarks, considering all three processes at two temperatures: T= 360 MeV (upper panel) and T= 480 MeV (lower panel). At high $p_{T}$, the $R_{AA}$ obtained due to the elastic collision is highly suppressed, whereas the same from the polarization process is least suppressed. However, the $R_{AA}$ due to the radiation loss is in between the two. If we move to a higher temperature, the $R_{AA}$ is found to be suppressed more in all three cases. In the intermediate range of $p_T \in (2,4)$GeV the $R_{AA}$ due to polarization case is most suppressed showing its important contribution in the process of HQ energy loss in the QGP medium. In this range, $R_{AA}$ due to elastic processes is in between the other two. 

{ Similarly, in Fig.~\ref{raa_b}, we show the plots of $R_{AA}$ for bottom quarks, taking all three processes at two temperatures: T= 360 MeV (upper panel) and T= 480 MeV (lower panel). At high $p_{T}$, the $R_{AA}$ is calculated for the elastic collision and the polarization process and found more suppressed (less value of $R_{AA}$). However, the $R_{AA}$ due to the radiation loss is least suppressed as it is consistent with Fig. \ref{drag_p} results.} Nevertheless, we conclude that considering the elastic effect along with the polarization effect and radiation effect is crucial for ensuring theoretical consistency in the characterization of the transport of the HQs in the QGP medium. Studying all three processes together might be helpful in the real extraction of the spatial diffusion coefficient of the HQ.

\section{Summary and Conclusions}
\label{SC}

There are several studies on the HQ analysis available in the literature where the main focus is only on the elastic and inelastic collision effects while ignoring the impact of polarization effects. However, recent studies have shown that elastic effects, along with polarization and inelastic effects, also play a significant role in the HQ analysis. In order to combine all three analyses together, this article studies the energy loss of the HQs considering these three processes. 
The results of the analysis show that the energy loss and drag coefficient are comparable for all three processes, specifically at low momentum. The $R_{AA}$ is calculated using the Langevin equation to refine the analysis further. The elastic effects are a major contribution compared to the polarization and radiation cases at high $p_T$, whereas the polarization contribution is substantial at low $p_T$. This suggests that accounting for elastic effects in conjunction with polarization and inelastic scattering processes is crucial in studying HQ propagation in the hot QGP medium, at least at low momentum. However, given the simplifying assumptions made in our present study, it is
difficult to predict the specific modifications. A more complete and quantitative study will be
conducted in future investigations to understand these phenomena further.

This analysis can be extended by considering the expanding QGP medium and the impact of polarization on the build-up of $v_2$ to understand experimental outcomes better. To do so, we need to include an exact initial geometry and an expanding QGP medium. Future extensions of the analysis can also incorporate memory effects. Overall, this study provides a more comprehensive understanding of the medium and its expansion in HIC experiments.

\section{Acknowledgements}
 We would like to acknowledge Dr. Santosh Kumar Das for the useful discussions. J.P. acknowledges the partial support from DAE-BRNS, India, Grant
No. 57/14/02/ 2021-BRNS also a partial support from the SERB Fellowship Project Code No.RD/0122-SERBF30-001. MY Jamal would like to acknowledge the SERB-NPDF (National postdoctoral fellow) with File No. PDF/2022/001551. \\
\vspace{3mm}

{\bf Data Availability Statement:} No Data associated in the manuscript

 \bibliography{ref1}

\begin{thebibliography}{73}%
\makeatletter
\providecommand \@ifxundefined [1]{%
 \@ifx{#1\undefined}
}%
\providecommand \@ifnum [1]{%
 \ifnum #1\expandafter \@firstoftwo
 \else \expandafter \@secondoftwo
 \fi
}%
\providecommand \@ifx [1]{%
 \ifx #1\expandafter \@firstoftwo
 \else \expandafter \@secondoftwo
 \fi
}%
\providecommand \natexlab [1]{#1}%
\providecommand \enquote  [1]{``#1''}%
\providecommand \bibnamefont  [1]{#1}%
\providecommand \bibfnamefont [1]{#1}%
\providecommand \citenamefont [1]{#1}%
\providecommand \href@noop [0]{\@secondoftwo}%
\providecommand \href [0]{\begingroup \@sanitize@url \@href}%
\providecommand \@href[1]{\@@startlink{#1}\@@href}%
\providecommand \@@href[1]{\endgroup#1\@@endlink}%
\providecommand \@sanitize@url [0]{\catcode `\\12\catcode `\$12\catcode
  `\&12\catcode `\#12\catcode `\^12\catcode `\_12\catcode `\%12\relax}%
\providecommand \@@startlink[1]{}%
\providecommand \@@endlink[0]{}%
\providecommand \url  [0]{\begingroup\@sanitize@url \@url }%
\providecommand \@url [1]{\endgroup\@href {#1}{\urlprefix }}%
\providecommand \urlprefix  [0]{URL }%
\providecommand \Eprint [0]{\href }%
\providecommand \doibase [0]{http://dx.doi.org/}%
\providecommand \selectlanguage [0]{\@gobble}%
\providecommand \bibinfo  [0]{\@secondoftwo}%
\providecommand \bibfield  [0]{\@secondoftwo}%
\providecommand \translation [1]{[#1]}%
\providecommand \BibitemOpen [0]{}%
\providecommand \bibitemStop [0]{}%
\providecommand \bibitemNoStop [0]{.\EOS\space}%
\providecommand \EOS [0]{\spacefactor3000\relax}%
\providecommand \BibitemShut  [1]{\csname bibitem#1\endcsname}%
\let\auto@bib@innerbib\@empty
\bibitem [{\citenamefont {Back}(2005)}]{PHOBOS:2004zne}%
  \BibitemOpen
  \bibfield  {author} {\bibinfo {author} {\bibfnamefont {B.~B. e.~a.}\
  \bibnamefont {Back}} (\bibinfo {collaboration} {PHOBOS}),\ }\href@noop {}
  {\bibfield  {journal} {\bibinfo  {journal} {Nucl. Phys. A}\ }\textbf
  {\bibinfo {volume} {757}},\ \bibinfo {pages} {28} (\bibinfo {year}
  {2005})}\BibitemShut {NoStop}%
\bibitem [{\citenamefont {Arsene}(2005)}]{BRAHMS:2004adc}%
  \BibitemOpen
  \bibfield  {author} {\bibinfo {author} {\bibfnamefont {I.~e.~a.}\
  \bibnamefont {Arsene}} (\bibinfo {collaboration} {BRAHMS}),\ }\href@noop {}
  {\bibfield  {journal} {\bibinfo  {journal} {Nucl. Phys. A}\ }\textbf
  {\bibinfo {volume} {757}},\ \bibinfo {pages} {1} (\bibinfo {year}
  {2005})}\BibitemShut {NoStop}%
\bibitem [{\citenamefont {Aamodt}\ \emph {et~al.}(2010)\citenamefont {Aamodt}
  \emph {et~al.}}]{ALICE:2010khr}%
  \BibitemOpen
  \bibfield  {author} {\bibinfo {author} {\bibfnamefont {K.}~\bibnamefont
  {Aamodt}} \emph {et~al.} (\bibinfo {collaboration} {ALICE}),\ }\href
  {\doibase 10.1103/PhysRevLett.105.252301} {\bibfield  {journal} {\bibinfo
  {journal} {Phys. Rev. Lett.}\ }\textbf {\bibinfo {volume} {105}},\ \bibinfo
  {pages} {252301} (\bibinfo {year} {2010})},\ \Eprint
  {http://arxiv.org/abs/1011.3916} {arXiv:1011.3916 [nucl-ex]} \BibitemShut
  {NoStop}%
\bibitem [{\citenamefont {Fukushima}\ \emph {et~al.}(2021)\citenamefont
  {Fukushima}, \citenamefont {Mohanty},\ and\ \citenamefont
  {Xu}}]{Fukushima:2020yzx}%
  \BibitemOpen
  \bibfield  {author} {\bibinfo {author} {\bibfnamefont {K.}~\bibnamefont
  {Fukushima}}, \bibinfo {author} {\bibfnamefont {B.}~\bibnamefont {Mohanty}},
  \ and\ \bibinfo {author} {\bibfnamefont {N.}~\bibnamefont {Xu}},\ }\href
  {\doibase 10.1007/s43673-021-00002-7} {\bibfield  {journal} {\bibinfo
  {journal} {AAPPS Bull.}\ }\textbf {\bibinfo {volume} {31}},\ \bibinfo {pages}
  {1} (\bibinfo {year} {2021})},\ \Eprint {http://arxiv.org/abs/2009.03006}
  {arXiv:2009.03006 [hep-ph]} \BibitemShut {NoStop}%
\bibitem [{\citenamefont {van Hees}\ \emph {et~al.}(2006)\citenamefont {van
  Hees}, \citenamefont {Greco},\ and\ \citenamefont {Rapp}}]{vanHees:2005wb}%
  \BibitemOpen
  \bibfield  {author} {\bibinfo {author} {\bibfnamefont {H.}~\bibnamefont {van
  Hees}}, \bibinfo {author} {\bibfnamefont {V.}~\bibnamefont {Greco}}, \ and\
  \bibinfo {author} {\bibfnamefont {R.}~\bibnamefont {Rapp}},\ }\href {\doibase
  10.1103/PhysRevC.73.034913} {\bibfield  {journal} {\bibinfo  {journal} {Phys.
  Rev. C}\ }\textbf {\bibinfo {volume} {73}},\ \bibinfo {pages} {034913}
  (\bibinfo {year} {2006})},\ \Eprint {http://arxiv.org/abs/nucl-th/0508055}
  {arXiv:nucl-th/0508055} \BibitemShut {NoStop}%
\bibitem [{\citenamefont {Das}\ \emph {et~al.}(2009)\citenamefont {Das},
  \citenamefont {Alam},\ and\ \citenamefont {Mohanty}}]{Das:2009vy}%
  \BibitemOpen
  \bibfield  {author} {\bibinfo {author} {\bibfnamefont {S.~K.}\ \bibnamefont
  {Das}}, \bibinfo {author} {\bibfnamefont {J.-e.}\ \bibnamefont {Alam}}, \
  and\ \bibinfo {author} {\bibfnamefont {P.}~\bibnamefont {Mohanty}},\ }\href
  {\doibase 10.1103/PhysRevC.80.054916} {\bibfield  {journal} {\bibinfo
  {journal} {Phys. Rev. C}\ }\textbf {\bibinfo {volume} {80}},\ \bibinfo
  {pages} {054916} (\bibinfo {year} {2009})},\ \Eprint
  {http://arxiv.org/abs/0908.4194} {arXiv:0908.4194 [nucl-th]} \BibitemShut
  {NoStop}%
\bibitem [{\citenamefont {Das}\ \emph {et~al.}(2014)\citenamefont {Das},
  \citenamefont {Scardina}, \citenamefont {Plumari},\ and\ \citenamefont
  {Greco}}]{Das:2013kea}%
  \BibitemOpen
  \bibfield  {author} {\bibinfo {author} {\bibfnamefont {S.~K.}\ \bibnamefont
  {Das}}, \bibinfo {author} {\bibfnamefont {F.}~\bibnamefont {Scardina}},
  \bibinfo {author} {\bibfnamefont {S.}~\bibnamefont {Plumari}}, \ and\
  \bibinfo {author} {\bibfnamefont {V.}~\bibnamefont {Greco}},\ }\href
  {\doibase 10.1103/PhysRevC.90.044901} {\bibfield  {journal} {\bibinfo
  {journal} {Phys. Rev. C}\ }\textbf {\bibinfo {volume} {90}},\ \bibinfo
  {pages} {044901} (\bibinfo {year} {2014})},\ \Eprint
  {http://arxiv.org/abs/1312.6857} {arXiv:1312.6857 [nucl-th]} \BibitemShut
  {NoStop}%
\bibitem [{\citenamefont {Cao}\ \emph {et~al.}(2019)\citenamefont {Cao} \emph
  {et~al.}}]{Cao:2018ews}%
  \BibitemOpen
  \bibfield  {author} {\bibinfo {author} {\bibfnamefont {S.}~\bibnamefont
  {Cao}} \emph {et~al.},\ }\href {\doibase 10.1103/PhysRevC.99.054907}
  {\bibfield  {journal} {\bibinfo  {journal} {Phys. Rev. C}\ }\textbf {\bibinfo
  {volume} {99}},\ \bibinfo {pages} {054907} (\bibinfo {year} {2019})},\
  \Eprint {http://arxiv.org/abs/1809.07894} {arXiv:1809.07894 [nucl-th]}
  \BibitemShut {NoStop}%
\bibitem [{\citenamefont {Song}\ \emph {et~al.}(2020)\citenamefont {Song},
  \citenamefont {Moreau}, \citenamefont {Xu}, \citenamefont {Ozvenchuk},
  \citenamefont {Bratkovskaya}, \citenamefont {Aichelin}, \citenamefont {Bass},
  \citenamefont {Gossiaux},\ and\ \citenamefont {Nahrgang}}]{Song:2020tfm}%
  \BibitemOpen
  \bibfield  {author} {\bibinfo {author} {\bibfnamefont {T.}~\bibnamefont
  {Song}}, \bibinfo {author} {\bibfnamefont {P.}~\bibnamefont {Moreau}},
  \bibinfo {author} {\bibfnamefont {Y.}~\bibnamefont {Xu}}, \bibinfo {author}
  {\bibfnamefont {V.}~\bibnamefont {Ozvenchuk}}, \bibinfo {author}
  {\bibfnamefont {E.}~\bibnamefont {Bratkovskaya}}, \bibinfo {author}
  {\bibfnamefont {J.}~\bibnamefont {Aichelin}}, \bibinfo {author}
  {\bibfnamefont {S.~A.}\ \bibnamefont {Bass}}, \bibinfo {author}
  {\bibfnamefont {P.~B.}\ \bibnamefont {Gossiaux}}, \ and\ \bibinfo {author}
  {\bibfnamefont {M.}~\bibnamefont {Nahrgang}},\ }\href {\doibase
  10.1103/PhysRevC.101.044903} {\bibfield  {journal} {\bibinfo  {journal}
  {Phys. Rev. C}\ }\textbf {\bibinfo {volume} {101}},\ \bibinfo {pages}
  {044903} (\bibinfo {year} {2020})},\ \Eprint
  {http://arxiv.org/abs/2001.07951} {arXiv:2001.07951 [nucl-th]} \BibitemShut
  {NoStop}%
\bibitem [{\citenamefont {Kurian}\ \emph {et~al.}(2020)\citenamefont {Kurian},
  \citenamefont {Singh}, \citenamefont {Chandra}, \citenamefont {Jeon},\ and\
  \citenamefont {Gale}}]{Kurian:2020orp}%
  \BibitemOpen
  \bibfield  {author} {\bibinfo {author} {\bibfnamefont {M.}~\bibnamefont
  {Kurian}}, \bibinfo {author} {\bibfnamefont {M.}~\bibnamefont {Singh}},
  \bibinfo {author} {\bibfnamefont {V.}~\bibnamefont {Chandra}}, \bibinfo
  {author} {\bibfnamefont {S.}~\bibnamefont {Jeon}}, \ and\ \bibinfo {author}
  {\bibfnamefont {C.}~\bibnamefont {Gale}},\ }\href {\doibase
  10.1103/PhysRevC.102.044907} {\bibfield  {journal} {\bibinfo  {journal}
  {Phys. Rev. C}\ }\textbf {\bibinfo {volume} {102}},\ \bibinfo {pages}
  {044907} (\bibinfo {year} {2020})},\ \Eprint
  {http://arxiv.org/abs/2007.07705} {arXiv:2007.07705 [hep-ph]} \BibitemShut
  {NoStop}%
\bibitem [{\citenamefont {Sebastian}\ \emph {et~al.}(2023)\citenamefont
  {Sebastian}, \citenamefont {Jamal},\ and\ \citenamefont
  {Haque}}]{Sebastian:2022sga}%
  \BibitemOpen
  \bibfield  {author} {\bibinfo {author} {\bibfnamefont {J.}~\bibnamefont
  {Sebastian}}, \bibinfo {author} {\bibfnamefont {M.~Y.}\ \bibnamefont
  {Jamal}}, \ and\ \bibinfo {author} {\bibfnamefont {N.}~\bibnamefont
  {Haque}},\ }\href {\doibase 10.1103/PhysRevD.107.054040} {\bibfield
  {journal} {\bibinfo  {journal} {Phys. Rev. D}\ }\textbf {\bibinfo {volume}
  {107}},\ \bibinfo {pages} {054040} (\bibinfo {year} {2023})},\ \Eprint
  {http://arxiv.org/abs/2207.08510} {arXiv:2207.08510 [hep-ph]} \BibitemShut
  {NoStop}%
\bibitem [{\citenamefont {Singh}\ \emph {et~al.}(2015)\citenamefont {Singh},
  \citenamefont {Srivastava}, \citenamefont {Ganesh},\ and\ \citenamefont
  {Mishra}}]{Singh:2015eta}%
  \BibitemOpen
  \bibfield  {author} {\bibinfo {author} {\bibfnamefont {C.~R.}\ \bibnamefont
  {Singh}}, \bibinfo {author} {\bibfnamefont {P.~K.}\ \bibnamefont
  {Srivastava}}, \bibinfo {author} {\bibfnamefont {S.}~\bibnamefont {Ganesh}},
  \ and\ \bibinfo {author} {\bibfnamefont {M.}~\bibnamefont {Mishra}},\ }\href
  {\doibase 10.1103/PhysRevC.92.034916} {\bibfield  {journal} {\bibinfo
  {journal} {Phys. Rev. C}\ }\textbf {\bibinfo {volume} {92}},\ \bibinfo
  {pages} {034916} (\bibinfo {year} {2015})},\ \Eprint
  {http://arxiv.org/abs/1505.05674} {arXiv:1505.05674 [hep-ph]} \BibitemShut
  {NoStop}%
\bibitem [{\citenamefont {Matsui}\ and\ \citenamefont
  {Satz}(1986)}]{Matsui:1986dk}%
  \BibitemOpen
  \bibfield  {author} {\bibinfo {author} {\bibfnamefont {T.}~\bibnamefont
  {Matsui}}\ and\ \bibinfo {author} {\bibfnamefont {H.}~\bibnamefont {Satz}},\
  }\href {\doibase 10.1016/0370-2693(86)91404-8} {\bibfield  {journal}
  {\bibinfo  {journal} {Phys. Lett. B}\ }\textbf {\bibinfo {volume} {178}},\
  \bibinfo {pages} {416} (\bibinfo {year} {1986})}\BibitemShut {NoStop}%
\bibitem [{\citenamefont {Jamal}\ \emph {et~al.}(2018)\citenamefont {Jamal},
  \citenamefont {Nilima}, \citenamefont {Chandra},\ and\ \citenamefont
  {Agotiya}}]{Jamal:2018mog}%
  \BibitemOpen
  \bibfield  {author} {\bibinfo {author} {\bibfnamefont {M.~Y.}\ \bibnamefont
  {Jamal}}, \bibinfo {author} {\bibfnamefont {I.}~\bibnamefont {Nilima}},
  \bibinfo {author} {\bibfnamefont {V.}~\bibnamefont {Chandra}}, \ and\
  \bibinfo {author} {\bibfnamefont {V.~K.}\ \bibnamefont {Agotiya}},\ }\href
  {\doibase 10.1103/PhysRevD.97.094033} {\bibfield  {journal} {\bibinfo
  {journal} {Phys. Rev. D}\ }\textbf {\bibinfo {volume} {97}},\ \bibinfo
  {pages} {094033} (\bibinfo {year} {2018})},\ \Eprint
  {http://arxiv.org/abs/1805.04763} {arXiv:1805.04763 [nucl-th]} \BibitemShut
  {NoStop}%
\bibitem [{\citenamefont {Agotiya}\ \emph {et~al.}(2016)\citenamefont
  {Agotiya}, \citenamefont {Chandra}, \citenamefont {Jamal},\ and\
  \citenamefont {Nilima}}]{Agotiya:2016bqr}%
  \BibitemOpen
  \bibfield  {author} {\bibinfo {author} {\bibfnamefont {V.~K.}\ \bibnamefont
  {Agotiya}}, \bibinfo {author} {\bibfnamefont {V.}~\bibnamefont {Chandra}},
  \bibinfo {author} {\bibfnamefont {M.~Y.}\ \bibnamefont {Jamal}}, \ and\
  \bibinfo {author} {\bibfnamefont {I.}~\bibnamefont {Nilima}},\ }\href
  {\doibase 10.1103/PhysRevD.94.094006} {\bibfield  {journal} {\bibinfo
  {journal} {Phys. Rev. D}\ }\textbf {\bibinfo {volume} {94}},\ \bibinfo
  {pages} {094006} (\bibinfo {year} {2016})},\ \Eprint
  {http://arxiv.org/abs/1610.03170} {arXiv:1610.03170 [nucl-th]} \BibitemShut
  {NoStop}%
\bibitem [{\citenamefont {Song}\ \emph {et~al.}(2015)\citenamefont {Song},
  \citenamefont {Berrehrah}, \citenamefont {Cabrera}, \citenamefont
  {Torres-Rincon}, \citenamefont {Tolos}, \citenamefont {Cassing},\ and\
  \citenamefont {Bratkovskaya}}]{Song:2015sfa}%
  \BibitemOpen
  \bibfield  {author} {\bibinfo {author} {\bibfnamefont {T.}~\bibnamefont
  {Song}}, \bibinfo {author} {\bibfnamefont {H.}~\bibnamefont {Berrehrah}},
  \bibinfo {author} {\bibfnamefont {D.}~\bibnamefont {Cabrera}}, \bibinfo
  {author} {\bibfnamefont {J.~M.}\ \bibnamefont {Torres-Rincon}}, \bibinfo
  {author} {\bibfnamefont {L.}~\bibnamefont {Tolos}}, \bibinfo {author}
  {\bibfnamefont {W.}~\bibnamefont {Cassing}}, \ and\ \bibinfo {author}
  {\bibfnamefont {E.}~\bibnamefont {Bratkovskaya}},\ }\href {\doibase
  10.1103/PhysRevC.92.014910} {\bibfield  {journal} {\bibinfo  {journal} {Phys.
  Rev. C}\ }\textbf {\bibinfo {volume} {92}},\ \bibinfo {pages} {014910}
  (\bibinfo {year} {2015})},\ \Eprint {http://arxiv.org/abs/1503.03039}
  {arXiv:1503.03039 [nucl-th]} \BibitemShut {NoStop}%
\bibitem [{\citenamefont {Beraudo}\ \emph {et~al.}(2018)\citenamefont {Beraudo}
  \emph {et~al.}}]{Rapp:2018qla}%
  \BibitemOpen
  \bibfield  {author} {\bibinfo {author} {\bibfnamefont {A.}~\bibnamefont
  {Beraudo}} \emph {et~al.},\ }\href {\doibase 10.1016/j.nuclphysa.2018.09.002}
  {\bibfield  {journal} {\bibinfo  {journal} {Nucl. Phys. A}\ }\textbf
  {\bibinfo {volume} {979}},\ \bibinfo {pages} {21} (\bibinfo {year} {2018})},\
  \Eprint {http://arxiv.org/abs/1803.03824} {arXiv:1803.03824 [nucl-th]}
  \BibitemShut {NoStop}%
\bibitem [{\citenamefont {Golam~Mustafa}\ \emph {et~al.}(1998)\citenamefont
  {Golam~Mustafa}, \citenamefont {Pal},\ and\ \citenamefont
  {Kumar~Srivastava}}]{GolamMustafa:1997id}%
  \BibitemOpen
  \bibfield  {author} {\bibinfo {author} {\bibfnamefont {M.}~\bibnamefont
  {Golam~Mustafa}}, \bibinfo {author} {\bibfnamefont {D.}~\bibnamefont {Pal}},
  \ and\ \bibinfo {author} {\bibfnamefont {D.}~\bibnamefont
  {Kumar~Srivastava}},\ }\href {\doibase 10.1103/PhysRevC.57.3499} {\bibfield
  {journal} {\bibinfo  {journal} {Phys. Rev. C}\ }\textbf {\bibinfo {volume}
  {57}},\ \bibinfo {pages} {889} (\bibinfo {year} {1998})},\ \bibinfo {note}
  {[Erratum: Phys.Rev.C 57, 3499--3499 (1998)]},\ \Eprint
  {http://arxiv.org/abs/nucl-th/9706001} {arXiv:nucl-th/9706001} \BibitemShut
  {NoStop}%
\bibitem [{\citenamefont {Plumari}\ \emph {et~al.}(2018)\citenamefont
  {Plumari}, \citenamefont {Minissale}, \citenamefont {Das}, \citenamefont
  {Coci},\ and\ \citenamefont {Greco}}]{Plumari:2017ntm}%
  \BibitemOpen
  \bibfield  {author} {\bibinfo {author} {\bibfnamefont {S.}~\bibnamefont
  {Plumari}}, \bibinfo {author} {\bibfnamefont {V.}~\bibnamefont {Minissale}},
  \bibinfo {author} {\bibfnamefont {S.~K.}\ \bibnamefont {Das}}, \bibinfo
  {author} {\bibfnamefont {G.}~\bibnamefont {Coci}}, \ and\ \bibinfo {author}
  {\bibfnamefont {V.}~\bibnamefont {Greco}},\ }\href {\doibase
  10.1140/epjc/s10052-018-5828-7} {\bibfield  {journal} {\bibinfo  {journal}
  {Eur. Phys. J. C}\ }\textbf {\bibinfo {volume} {78}},\ \bibinfo {pages} {348}
  (\bibinfo {year} {2018})},\ \Eprint {http://arxiv.org/abs/1712.00730}
  {arXiv:1712.00730 [hep-ph]} \BibitemShut {NoStop}%
\bibitem [{\citenamefont {Gossiaux}\ and\ \citenamefont
  {Aichelin}(2008)}]{Gossiaux:2008jv}%
  \BibitemOpen
  \bibfield  {author} {\bibinfo {author} {\bibfnamefont {P.~B.}\ \bibnamefont
  {Gossiaux}}\ and\ \bibinfo {author} {\bibfnamefont {J.}~\bibnamefont
  {Aichelin}},\ }\href {\doibase 10.1103/PhysRevC.78.014904} {\bibfield
  {journal} {\bibinfo  {journal} {Phys. Rev. C}\ }\textbf {\bibinfo {volume}
  {78}},\ \bibinfo {pages} {014904} (\bibinfo {year} {2008})},\ \Eprint
  {http://arxiv.org/abs/0802.2525} {arXiv:0802.2525 [hep-ph]} \BibitemShut
  {NoStop}%
\bibitem [{\citenamefont {Prakash}\ \emph {et~al.}(2021)\citenamefont
  {Prakash}, \citenamefont {Kurian}, \citenamefont {Das},\ and\ \citenamefont
  {Chandra}}]{Prakash:2021lwt}%
  \BibitemOpen
  \bibfield  {author} {\bibinfo {author} {\bibfnamefont {J.}~\bibnamefont
  {Prakash}}, \bibinfo {author} {\bibfnamefont {M.}~\bibnamefont {Kurian}},
  \bibinfo {author} {\bibfnamefont {S.~K.}\ \bibnamefont {Das}}, \ and\
  \bibinfo {author} {\bibfnamefont {V.}~\bibnamefont {Chandra}},\ }\href
  {\doibase 10.1103/PhysRevD.103.094009} {\bibfield  {journal} {\bibinfo
  {journal} {Phys. Rev. D}\ }\textbf {\bibinfo {volume} {103}},\ \bibinfo
  {pages} {094009} (\bibinfo {year} {2021})},\ \Eprint
  {http://arxiv.org/abs/2102.07082} {arXiv:2102.07082 [hep-ph]} \BibitemShut
  {NoStop}%
\bibitem [{\citenamefont {Prakash}\ \emph {et~al.}(2023)\citenamefont
  {Prakash}, \citenamefont {Chandra},\ and\ \citenamefont
  {Das}}]{Prakash:2023wbs}%
  \BibitemOpen
  \bibfield  {author} {\bibinfo {author} {\bibfnamefont {J.}~\bibnamefont
  {Prakash}}, \bibinfo {author} {\bibfnamefont {V.}~\bibnamefont {Chandra}}, \
  and\ \bibinfo {author} {\bibfnamefont {S.~K.}\ \bibnamefont {Das}},\ }\href
  {\doibase 10.1103/PhysRevD.108.096016} {\bibfield  {journal} {\bibinfo
  {journal} {Phys. Rev. D}\ }\textbf {\bibinfo {volume} {108}},\ \bibinfo
  {pages} {096016} (\bibinfo {year} {2023})},\ \Eprint
  {http://arxiv.org/abs/2306.07966} {arXiv:2306.07966 [hep-ph]} \BibitemShut
  {NoStop}%
\bibitem [{\citenamefont {Jamal}\ \emph {et~al.}(2024)\citenamefont {Jamal},
  \citenamefont {Prakash}, \citenamefont {Nilima},\ and\ \citenamefont
  {Bandyopadhyay}}]{Jamal:2023ncn}%
  \BibitemOpen
  \bibfield  {author} {\bibinfo {author} {\bibfnamefont {M.~Y.}\ \bibnamefont
  {Jamal}}, \bibinfo {author} {\bibfnamefont {J.}~\bibnamefont {Prakash}},
  \bibinfo {author} {\bibfnamefont {I.}~\bibnamefont {Nilima}}, \ and\ \bibinfo
  {author} {\bibfnamefont {A.}~\bibnamefont {Bandyopadhyay}},\ }\href {\doibase
  10.1088/1361-6471/ad290d} {\bibfield  {journal} {\bibinfo  {journal} {J.
  Phys. G}\ }\textbf {\bibinfo {volume} {51}},\ \bibinfo {pages} {045104}
  (\bibinfo {year} {2024})},\ \Eprint {http://arxiv.org/abs/2304.09851}
  {arXiv:2304.09851 [hep-ph]} \BibitemShut {NoStop}%
\bibitem [{\citenamefont {Singh}\ \emph {et~al.}(2023)\citenamefont {Singh},
  \citenamefont {Kurian}, \citenamefont {Jeon},\ and\ \citenamefont
  {Gale}}]{Singh:2023smw}%
  \BibitemOpen
  \bibfield  {author} {\bibinfo {author} {\bibfnamefont {M.}~\bibnamefont
  {Singh}}, \bibinfo {author} {\bibfnamefont {M.}~\bibnamefont {Kurian}},
  \bibinfo {author} {\bibfnamefont {S.}~\bibnamefont {Jeon}}, \ and\ \bibinfo
  {author} {\bibfnamefont {C.}~\bibnamefont {Gale}},\ }\href {\doibase
  10.1103/PhysRevC.108.054901} {\bibfield  {journal} {\bibinfo  {journal}
  {Phys. Rev. C}\ }\textbf {\bibinfo {volume} {108}},\ \bibinfo {pages}
  {054901} (\bibinfo {year} {2023})},\ \Eprint
  {http://arxiv.org/abs/2306.09514} {arXiv:2306.09514 [nucl-th]} \BibitemShut
  {NoStop}%
\bibitem [{\citenamefont {Cao}\ \emph {et~al.}(2016)\citenamefont {Cao},
  \citenamefont {Luo}, \citenamefont {Qin},\ and\ \citenamefont
  {Wang}}]{Cao:2016gvr}%
  \BibitemOpen
  \bibfield  {author} {\bibinfo {author} {\bibfnamefont {S.}~\bibnamefont
  {Cao}}, \bibinfo {author} {\bibfnamefont {T.}~\bibnamefont {Luo}}, \bibinfo
  {author} {\bibfnamefont {G.-Y.}\ \bibnamefont {Qin}}, \ and\ \bibinfo
  {author} {\bibfnamefont {X.-N.}\ \bibnamefont {Wang}},\ }\href {\doibase
  10.1103/PhysRevC.94.014909} {\bibfield  {journal} {\bibinfo  {journal} {Phys.
  Rev. C}\ }\textbf {\bibinfo {volume} {94}},\ \bibinfo {pages} {014909}
  (\bibinfo {year} {2016})},\ \Eprint {http://arxiv.org/abs/1605.06447}
  {arXiv:1605.06447 [nucl-th]} \BibitemShut {NoStop}%
\bibitem [{\citenamefont {Mazumder}\ \emph {et~al.}(2011)\citenamefont
  {Mazumder}, \citenamefont {Bhattacharyya}, \citenamefont {Alam},\ and\
  \citenamefont {Das}}]{Mazumder:2011nj}%
  \BibitemOpen
  \bibfield  {author} {\bibinfo {author} {\bibfnamefont {S.}~\bibnamefont
  {Mazumder}}, \bibinfo {author} {\bibfnamefont {T.}~\bibnamefont
  {Bhattacharyya}}, \bibinfo {author} {\bibfnamefont {J.-e.}\ \bibnamefont
  {Alam}}, \ and\ \bibinfo {author} {\bibfnamefont {S.~K.}\ \bibnamefont
  {Das}},\ }\href {\doibase 10.1103/PhysRevC.84.044901} {\bibfield  {journal}
  {\bibinfo  {journal} {Phys. Rev. C}\ }\textbf {\bibinfo {volume} {84}},\
  \bibinfo {pages} {044901} (\bibinfo {year} {2011})},\ \Eprint
  {http://arxiv.org/abs/1106.2615} {arXiv:1106.2615 [nucl-th]} \BibitemShut
  {NoStop}%
\bibitem [{\citenamefont {Zhang}\ \emph {et~al.}(2023)\citenamefont {Zhang},
  \citenamefont {Zheng}, \citenamefont {Shi},\ and\ \citenamefont
  {Lin}}]{Zhang:2022fum}%
  \BibitemOpen
  \bibfield  {author} {\bibinfo {author} {\bibfnamefont {C.}~\bibnamefont
  {Zhang}}, \bibinfo {author} {\bibfnamefont {L.}~\bibnamefont {Zheng}},
  \bibinfo {author} {\bibfnamefont {S.}~\bibnamefont {Shi}}, \ and\ \bibinfo
  {author} {\bibfnamefont {Z.-W.}\ \bibnamefont {Lin}},\ }\href {\doibase
  10.1016/j.physletb.2023.138219} {\bibfield  {journal} {\bibinfo  {journal}
  {Phys. Lett. B}\ }\textbf {\bibinfo {volume} {846}},\ \bibinfo {pages}
  {138219} (\bibinfo {year} {2023})},\ \Eprint
  {http://arxiv.org/abs/2210.07767} {arXiv:2210.07767 [nucl-th]} \BibitemShut
  {NoStop}%
\bibitem [{\citenamefont {Jamal}\ \emph
  {et~al.}(2021{\natexlab{a}})\citenamefont {Jamal}, \citenamefont {Das},\ and\
  \citenamefont {Ruggieri}}]{PhysRevD.103.054030}%
  \BibitemOpen
  \bibfield  {author} {\bibinfo {author} {\bibfnamefont {M.~Y.}\ \bibnamefont
  {Jamal}}, \bibinfo {author} {\bibfnamefont {S.~K.}\ \bibnamefont {Das}}, \
  and\ \bibinfo {author} {\bibfnamefont {M.}~\bibnamefont {Ruggieri}},\ }\href
  {\doibase 10.1103/PhysRevD.103.054030} {\bibfield  {journal} {\bibinfo
  {journal} {Phys. Rev. D}\ }\textbf {\bibinfo {volume} {103}},\ \bibinfo
  {pages} {054030} (\bibinfo {year} {2021}{\natexlab{a}})}\BibitemShut
  {NoStop}%
\bibitem [{\citenamefont {Jamal}\ and\ \citenamefont
  {Mohanty}(2021{\natexlab{a}})}]{Jamal:2021btg}%
  \BibitemOpen
  \bibfield  {author} {\bibinfo {author} {\bibfnamefont {M.~Y.}\ \bibnamefont
  {Jamal}}\ and\ \bibinfo {author} {\bibfnamefont {B.}~\bibnamefont
  {Mohanty}},\ }\href {\doibase 10.1140/epjc/s10052-021-09418-9} {\bibfield
  {journal} {\bibinfo  {journal} {Eur. Phys. J. C}\ }\textbf {\bibinfo {volume}
  {81}},\ \bibinfo {pages} {616} (\bibinfo {year} {2021}{\natexlab{a}})},\
  \Eprint {http://arxiv.org/abs/2101.00164} {arXiv:2101.00164 [nucl-th]}
  \BibitemShut {NoStop}%
\bibitem [{\citenamefont {Jamal}\ and\ \citenamefont
  {Mohanty}(2021{\natexlab{b}})}]{Jamal:2020emj}%
  \BibitemOpen
  \bibfield  {author} {\bibinfo {author} {\bibfnamefont {M.~Y.}\ \bibnamefont
  {Jamal}}\ and\ \bibinfo {author} {\bibfnamefont {B.}~\bibnamefont
  {Mohanty}},\ }\href {\doibase 10.1140/epjp/s13360-021-01098-4} {\bibfield
  {journal} {\bibinfo  {journal} {Eur. Phys. J. Plus}\ }\textbf {\bibinfo
  {volume} {136}},\ \bibinfo {pages} {130} (\bibinfo {year}
  {2021}{\natexlab{b}})},\ \Eprint {http://arxiv.org/abs/2002.09230}
  {arXiv:2002.09230 [nucl-th]} \BibitemShut {NoStop}%
\bibitem [{\citenamefont {Sun}\ \emph {et~al.}(2023)\citenamefont {Sun},
  \citenamefont {Plumari},\ and\ \citenamefont {Das}}]{Sun:2023adv}%
  \BibitemOpen
  \bibfield  {author} {\bibinfo {author} {\bibfnamefont {Y.}~\bibnamefont
  {Sun}}, \bibinfo {author} {\bibfnamefont {S.}~\bibnamefont {Plumari}}, \ and\
  \bibinfo {author} {\bibfnamefont {S.~K.}\ \bibnamefont {Das}},\ }\href
  {\doibase 10.1016/j.physletb.2023.138043} {\bibfield  {journal} {\bibinfo
  {journal} {Phys. Lett. B}\ }\textbf {\bibinfo {volume} {843}},\ \bibinfo
  {pages} {138043} (\bibinfo {year} {2023})},\ \Eprint
  {http://arxiv.org/abs/2304.12792} {arXiv:2304.12792 [nucl-th]} \BibitemShut
  {NoStop}%
\bibitem [{\citenamefont {Plumari}\ \emph {et~al.}(2020)\citenamefont
  {Plumari}, \citenamefont {Coci}, \citenamefont {Minissale}, \citenamefont
  {Das}, \citenamefont {Sun},\ and\ \citenamefont {Greco}}]{Plumari:2019hzp}%
  \BibitemOpen
  \bibfield  {author} {\bibinfo {author} {\bibfnamefont {S.}~\bibnamefont
  {Plumari}}, \bibinfo {author} {\bibfnamefont {G.}~\bibnamefont {Coci}},
  \bibinfo {author} {\bibfnamefont {V.}~\bibnamefont {Minissale}}, \bibinfo
  {author} {\bibfnamefont {S.~K.}\ \bibnamefont {Das}}, \bibinfo {author}
  {\bibfnamefont {Y.}~\bibnamefont {Sun}}, \ and\ \bibinfo {author}
  {\bibfnamefont {V.}~\bibnamefont {Greco}},\ }\href {\doibase
  10.1016/j.physletb.2020.135460} {\bibfield  {journal} {\bibinfo  {journal}
  {Phys. Lett. B}\ }\textbf {\bibinfo {volume} {805}},\ \bibinfo {pages}
  {135460} (\bibinfo {year} {2020})},\ \Eprint
  {http://arxiv.org/abs/1912.09350} {arXiv:1912.09350 [hep-ph]} \BibitemShut
  {NoStop}%
\bibitem [{\citenamefont {Du}\ and\ \citenamefont
  {Qian}(2023)}]{du2023accelerated}%
  \BibitemOpen
  \bibfield  {author} {\bibinfo {author} {\bibfnamefont {X.}~\bibnamefont
  {Du}}\ and\ \bibinfo {author} {\bibfnamefont {W.}~\bibnamefont {Qian}},\
  }\href@noop {} {\bibfield  {journal} {\bibinfo  {journal} {arXiv preprint
  arXiv:2312.16294}\ } (\bibinfo {year} {2023})}\BibitemShut {NoStop}%
\bibitem [{\citenamefont {Shaikh}\ \emph {et~al.}(2021)\citenamefont {Shaikh},
  \citenamefont {Kurian}, \citenamefont {Das}, \citenamefont {Chandra},
  \citenamefont {Dash},\ and\ \citenamefont {Nandi}}]{Shaikh:2021lka}%
  \BibitemOpen
  \bibfield  {author} {\bibinfo {author} {\bibfnamefont {A.}~\bibnamefont
  {Shaikh}}, \bibinfo {author} {\bibfnamefont {M.}~\bibnamefont {Kurian}},
  \bibinfo {author} {\bibfnamefont {S.~K.}\ \bibnamefont {Das}}, \bibinfo
  {author} {\bibfnamefont {V.}~\bibnamefont {Chandra}}, \bibinfo {author}
  {\bibfnamefont {S.}~\bibnamefont {Dash}}, \ and\ \bibinfo {author}
  {\bibfnamefont {B.~K.}\ \bibnamefont {Nandi}},\ }\href {\doibase
  10.1103/PhysRevD.104.034017} {\bibfield  {journal} {\bibinfo  {journal}
  {Phys. Rev. D}\ }\textbf {\bibinfo {volume} {104}},\ \bibinfo {pages}
  {034017} (\bibinfo {year} {2021})},\ \Eprint
  {http://arxiv.org/abs/2105.14296} {arXiv:2105.14296 [hep-ph]} \BibitemShut
  {NoStop}%
\bibitem [{\citenamefont {Kumar}\ \emph {et~al.}(2022)\citenamefont {Kumar},
  \citenamefont {Kurian}, \citenamefont {Das},\ and\ \citenamefont
  {Chandra}}]{Kumar:2021goi}%
  \BibitemOpen
  \bibfield  {author} {\bibinfo {author} {\bibfnamefont {A.}~\bibnamefont
  {Kumar}}, \bibinfo {author} {\bibfnamefont {M.}~\bibnamefont {Kurian}},
  \bibinfo {author} {\bibfnamefont {S.~K.}\ \bibnamefont {Das}}, \ and\
  \bibinfo {author} {\bibfnamefont {V.}~\bibnamefont {Chandra}},\ }\href
  {\doibase 10.1103/PhysRevC.105.054903} {\bibfield  {journal} {\bibinfo
  {journal} {Phys. Rev. C}\ }\textbf {\bibinfo {volume} {105}},\ \bibinfo
  {pages} {054903} (\bibinfo {year} {2022})},\ \Eprint
  {http://arxiv.org/abs/2111.07563} {arXiv:2111.07563 [hep-ph]} \BibitemShut
  {NoStop}%
\bibitem [{\citenamefont {Das}\ \emph {et~al.}(2022)\citenamefont {Das} \emph
  {et~al.}}]{Das:2022lqh}%
  \BibitemOpen
  \bibfield  {author} {\bibinfo {author} {\bibfnamefont {S.~K.}\ \bibnamefont
  {Das}} \emph {et~al.},\ }\href {\doibase 10.1142/S0218301322500975}
  {\bibfield  {journal} {\bibinfo  {journal} {Int. J. Mod. Phys. E}\ }\textbf
  {\bibinfo {volume} {31}},\ \bibinfo {pages} {12} (\bibinfo {year} {2022})},\
  \Eprint {http://arxiv.org/abs/2208.13440} {arXiv:2208.13440 [nucl-th]}
  \BibitemShut {NoStop}%
\bibitem [{\citenamefont {Bandyopadhyay}(2024)}]{Bandyopadhyay:2023hiv}%
  \BibitemOpen
  \bibfield  {author} {\bibinfo {author} {\bibfnamefont {A.}~\bibnamefont
  {Bandyopadhyay}},\ }\href {\doibase 10.1103/PhysRevD.109.034013} {\bibfield
  {journal} {\bibinfo  {journal} {Phys. Rev. D}\ }\textbf {\bibinfo {volume}
  {109}},\ \bibinfo {pages} {034013} (\bibinfo {year} {2024})},\ \Eprint
  {http://arxiv.org/abs/2307.09655} {arXiv:2307.09655 [hep-ph]} \BibitemShut
  {NoStop}%
\bibitem [{\citenamefont {Romatschke}\ and\ \citenamefont
  {Romatschke}(2007)}]{Romatschke:2007mq}%
  \BibitemOpen
  \bibfield  {author} {\bibinfo {author} {\bibfnamefont {P.}~\bibnamefont
  {Romatschke}}\ and\ \bibinfo {author} {\bibfnamefont {U.}~\bibnamefont
  {Romatschke}},\ }\href {\doibase 10.1103/PhysRevLett.99.172301} {\bibfield
  {journal} {\bibinfo  {journal} {Phys. Rev. Lett.}\ }\textbf {\bibinfo
  {volume} {99}},\ \bibinfo {pages} {172301} (\bibinfo {year} {2007})},\
  \Eprint {http://arxiv.org/abs/0706.1522} {arXiv:0706.1522 [nucl-th]}
  \BibitemShut {NoStop}%
\bibitem [{\citenamefont {Ryu}\ \emph {et~al.}(2015)\citenamefont {Ryu},
  \citenamefont {Paquet}, \citenamefont {Shen}, \citenamefont {Denicol},
  \citenamefont {Schenke}, \citenamefont {Jeon},\ and\ \citenamefont
  {Gale}}]{Ryu:2015vwa}%
  \BibitemOpen
  \bibfield  {author} {\bibinfo {author} {\bibfnamefont {S.}~\bibnamefont
  {Ryu}}, \bibinfo {author} {\bibfnamefont {J.~F.}\ \bibnamefont {Paquet}},
  \bibinfo {author} {\bibfnamefont {C.}~\bibnamefont {Shen}}, \bibinfo {author}
  {\bibfnamefont {G.~S.}\ \bibnamefont {Denicol}}, \bibinfo {author}
  {\bibfnamefont {B.}~\bibnamefont {Schenke}}, \bibinfo {author} {\bibfnamefont
  {S.}~\bibnamefont {Jeon}}, \ and\ \bibinfo {author} {\bibfnamefont
  {C.}~\bibnamefont {Gale}},\ }\href {\doibase 10.1103/PhysRevLett.115.132301}
  {\bibfield  {journal} {\bibinfo  {journal} {Phys. Rev. Lett.}\ }\textbf
  {\bibinfo {volume} {115}},\ \bibinfo {pages} {132301} (\bibinfo {year}
  {2015})},\ \Eprint {http://arxiv.org/abs/1502.01675} {arXiv:1502.01675
  [nucl-th]} \BibitemShut {NoStop}%
\bibitem [{\citenamefont {Prakash}(2024{\natexlab{a}})}]{Prakash:2024rdz}%
  \BibitemOpen
  \bibfield  {author} {\bibinfo {author} {\bibfnamefont {J.}~\bibnamefont
  {Prakash}},\ }\href {\doibase 10.1103/PhysRevD.109.114004} {\bibfield
  {journal} {\bibinfo  {journal} {Phys. Rev. D}\ }\textbf {\bibinfo {volume}
  {109}},\ \bibinfo {pages} {114004} (\bibinfo {year} {2024}{\natexlab{a}})},\
  \Eprint {http://arxiv.org/abs/2401.03757} {arXiv:2401.03757 [hep-ph]}
  \BibitemShut {NoStop}%
\bibitem [{\citenamefont {Ruggieri}\ \emph {et~al.}(2022)\citenamefont
  {Ruggieri}, \citenamefont {Pooja}, \citenamefont {Prakash},\ and\
  \citenamefont {Das}}]{Ruggieri:2022kxv}%
  \BibitemOpen
  \bibfield  {author} {\bibinfo {author} {\bibfnamefont {M.}~\bibnamefont
  {Ruggieri}}, \bibinfo {author} {\bibnamefont {Pooja}}, \bibinfo {author}
  {\bibfnamefont {J.}~\bibnamefont {Prakash}}, \ and\ \bibinfo {author}
  {\bibfnamefont {S.~K.}\ \bibnamefont {Das}},\ }\href {\doibase
  10.1103/PhysRevD.106.034032} {\bibfield  {journal} {\bibinfo  {journal}
  {Phys. Rev. D}\ }\textbf {\bibinfo {volume} {106}},\ \bibinfo {pages}
  {034032} (\bibinfo {year} {2022})},\ \Eprint
  {http://arxiv.org/abs/2203.06712} {arXiv:2203.06712 [hep-ph]} \BibitemShut
  {NoStop}%
\bibitem [{\citenamefont {Sumit}\ \emph {et~al.}(2024)\citenamefont {Sumit},
  \citenamefont {Mukherjee}, \citenamefont {Haque},\ and\ \citenamefont
  {Patra}}]{Sumit:2023oib}%
  \BibitemOpen
  \bibfield  {author} {\bibinfo {author} {\bibnamefont {Sumit}}, \bibinfo
  {author} {\bibfnamefont {A.}~\bibnamefont {Mukherjee}}, \bibinfo {author}
  {\bibfnamefont {N.}~\bibnamefont {Haque}}, \ and\ \bibinfo {author}
  {\bibfnamefont {B.~K.}\ \bibnamefont {Patra}},\ }\href {\doibase
  10.1103/PhysRevD.109.114043} {\bibfield  {journal} {\bibinfo  {journal}
  {Phys. Rev. D}\ }\textbf {\bibinfo {volume} {109}},\ \bibinfo {pages}
  {114043} (\bibinfo {year} {2024})},\ \Eprint
  {http://arxiv.org/abs/2311.18560} {arXiv:2311.18560 [hep-ph]} \BibitemShut
  {NoStop}%
\bibitem [{\citenamefont {Du}\ and\ \citenamefont {Qian}(2024)}]{Du:2023ewh}%
  \BibitemOpen
  \bibfield  {author} {\bibinfo {author} {\bibfnamefont {X.}~\bibnamefont
  {Du}}\ and\ \bibinfo {author} {\bibfnamefont {W.}~\bibnamefont {Qian}},\
  }\href {\doibase 10.1103/PhysRevD.109.076025} {\bibfield  {journal} {\bibinfo
   {journal} {Phys. Rev. D}\ }\textbf {\bibinfo {volume} {109}},\ \bibinfo
  {pages} {076025} (\bibinfo {year} {2024})},\ \Eprint
  {http://arxiv.org/abs/2312.16294} {arXiv:2312.16294 [hep-ph]} \BibitemShut
  {NoStop}%
\bibitem [{\citenamefont {Prakash}(2024{\natexlab{b}})}]{Prakash:2024irm}%
  \BibitemOpen
  \bibfield  {author} {\bibinfo {author} {\bibfnamefont {J.}~\bibnamefont
  {Prakash}},\ }\href@noop {} {\  (\bibinfo {year} {2024}{\natexlab{b}})},\
  \Eprint {http://arxiv.org/abs/2406.18714} {arXiv:2406.18714 [hep-ph]}
  \BibitemShut {NoStop}%
\bibitem [{\citenamefont {Carrington}\ \emph {et~al.}(2015)\citenamefont
  {Carrington}, \citenamefont {Deja},\ and\ \citenamefont
  {Mrowczynski}}]{Carrington:2015xca}%
  \BibitemOpen
  \bibfield  {author} {\bibinfo {author} {\bibfnamefont {M.~E.}\ \bibnamefont
  {Carrington}}, \bibinfo {author} {\bibfnamefont {K.}~\bibnamefont {Deja}}, \
  and\ \bibinfo {author} {\bibfnamefont {S.}~\bibnamefont {Mrowczynski}},\
  }\href {\doibase 10.1103/PhysRevC.92.044914} {\bibfield  {journal} {\bibinfo
  {journal} {Phys. Rev. C}\ }\textbf {\bibinfo {volume} {92}},\ \bibinfo
  {pages} {044914} (\bibinfo {year} {2015})},\ \Eprint
  {http://arxiv.org/abs/1506.09082} {arXiv:1506.09082 [hep-ph]} \BibitemShut
  {NoStop}%
\bibitem [{\citenamefont {Carrington}\ \emph {et~al.}(2017)\citenamefont
  {Carrington}, \citenamefont {Mr\'owczy\'nski},\ and\ \citenamefont
  {Schenke}}]{Carrington:2016mhd}%
  \BibitemOpen
  \bibfield  {author} {\bibinfo {author} {\bibfnamefont {M.~E.}\ \bibnamefont
  {Carrington}}, \bibinfo {author} {\bibfnamefont {S.}~\bibnamefont
  {Mr\'owczy\'nski}}, \ and\ \bibinfo {author} {\bibfnamefont {B.}~\bibnamefont
  {Schenke}},\ }\href {\doibase 10.1103/PhysRevC.95.024906} {\bibfield
  {journal} {\bibinfo  {journal} {Phys. Rev. C}\ }\textbf {\bibinfo {volume}
  {95}},\ \bibinfo {pages} {024906} (\bibinfo {year} {2017})},\ \Eprint
  {http://arxiv.org/abs/1607.02359} {arXiv:1607.02359 [hep-ph]} \BibitemShut
  {NoStop}%
\bibitem [{\citenamefont {Jamal}\ \emph
  {et~al.}(2021{\natexlab{b}})\citenamefont {Jamal}, \citenamefont {Das},\ and\
  \citenamefont {Ruggieri}}]{Jamal:2020fxo}%
  \BibitemOpen
  \bibfield  {author} {\bibinfo {author} {\bibfnamefont {M.~Y.}\ \bibnamefont
  {Jamal}}, \bibinfo {author} {\bibfnamefont {S.~K.}\ \bibnamefont {Das}}, \
  and\ \bibinfo {author} {\bibfnamefont {M.}~\bibnamefont {Ruggieri}},\ }\href
  {\doibase 10.1103/PhysRevD.103.054030} {\bibfield  {journal} {\bibinfo
  {journal} {Phys. Rev. D}\ }\textbf {\bibinfo {volume} {103}},\ \bibinfo
  {pages} {054030} (\bibinfo {year} {2021}{\natexlab{b}})},\ \Eprint
  {http://arxiv.org/abs/2009.00561} {arXiv:2009.00561 [nucl-th]} \BibitemShut
  {NoStop}%
\bibitem [{\citenamefont {Yousuf~Jamal}\ and\ \citenamefont
  {Chandra}(2019)}]{YousufJamal:2019pen}%
  \BibitemOpen
  \bibfield  {author} {\bibinfo {author} {\bibfnamefont {M.}~\bibnamefont
  {Yousuf~Jamal}}\ and\ \bibinfo {author} {\bibfnamefont {V.}~\bibnamefont
  {Chandra}},\ }\href {\doibase 10.1140/epjc/s10052-019-7278-2} {\bibfield
  {journal} {\bibinfo  {journal} {Eur. Phys. J. C}\ }\textbf {\bibinfo {volume}
  {79}},\ \bibinfo {pages} {761} (\bibinfo {year} {2019})},\ \Eprint
  {http://arxiv.org/abs/1907.12033} {arXiv:1907.12033 [nucl-th]} \BibitemShut
  {NoStop}%
\bibitem [{\citenamefont {Chakraborty}\ \emph {et~al.}(2007)\citenamefont
  {Chakraborty}, \citenamefont {Mustafa},\ and\ \citenamefont
  {Thoma}}]{Chakraborty:2006db}%
  \BibitemOpen
  \bibfield  {author} {\bibinfo {author} {\bibfnamefont {P.}~\bibnamefont
  {Chakraborty}}, \bibinfo {author} {\bibfnamefont {M.~G.}\ \bibnamefont
  {Mustafa}}, \ and\ \bibinfo {author} {\bibfnamefont {M.~H.}\ \bibnamefont
  {Thoma}},\ }\href {\doibase 10.1103/PhysRevC.75.064908} {\bibfield  {journal}
  {\bibinfo  {journal} {Phys. Rev. C}\ }\textbf {\bibinfo {volume} {75}},\
  \bibinfo {pages} {064908} (\bibinfo {year} {2007})},\ \Eprint
  {http://arxiv.org/abs/hep-ph/0611355} {arXiv:hep-ph/0611355} \BibitemShut
  {NoStop}%
\bibitem [{\citenamefont {Ghosh}\ \emph {et~al.}(2023)\citenamefont {Ghosh},
  \citenamefont {Jamal},\ and\ \citenamefont {Kurian}}]{Ghosh:2023ghi}%
  \BibitemOpen
  \bibfield  {author} {\bibinfo {author} {\bibfnamefont {R.}~\bibnamefont
  {Ghosh}}, \bibinfo {author} {\bibfnamefont {M.~Y.}\ \bibnamefont {Jamal}}, \
  and\ \bibinfo {author} {\bibfnamefont {M.}~\bibnamefont {Kurian}},\ }\href
  {\doibase 10.1103/PhysRevD.108.054035} {\bibfield  {journal} {\bibinfo
  {journal} {Phys. Rev. D}\ }\textbf {\bibinfo {volume} {108}},\ \bibinfo
  {pages} {054035} (\bibinfo {year} {2023})},\ \Eprint
  {http://arxiv.org/abs/2306.10247} {arXiv:2306.10247 [hep-ph]} \BibitemShut
  {NoStop}%
\bibitem [{\citenamefont {K}\ and\ \citenamefont {Chandra}(2023)}]{K:2023dum}%
  \BibitemOpen
  \bibfield  {author} {\bibinfo {author} {\bibfnamefont {G.~K.}\ \bibnamefont
  {K}}\ and\ \bibinfo {author} {\bibfnamefont {V.}~\bibnamefont {Chandra}},\
  }\href {\doibase 10.1103/PhysRevD.108.114015} {\bibfield  {journal} {\bibinfo
   {journal} {Phys. Rev. D}\ }\textbf {\bibinfo {volume} {108}},\ \bibinfo
  {pages} {114015} (\bibinfo {year} {2023})},\ \Eprint
  {http://arxiv.org/abs/2309.07118} {arXiv:2309.07118 [hep-ph]} \BibitemShut
  {NoStop}%
\bibitem [{\citenamefont {Debnath}\ \emph {et~al.}(2024)\citenamefont
  {Debnath}, \citenamefont {Ghosh}, \citenamefont {Jamal}, \citenamefont
  {Kurian},\ and\ \citenamefont {Prakash}}]{Debnath:2023zet}%
  \BibitemOpen
  \bibfield  {author} {\bibinfo {author} {\bibfnamefont {M.}~\bibnamefont
  {Debnath}}, \bibinfo {author} {\bibfnamefont {R.}~\bibnamefont {Ghosh}},
  \bibinfo {author} {\bibfnamefont {M.~Y.}\ \bibnamefont {Jamal}}, \bibinfo
  {author} {\bibfnamefont {M.}~\bibnamefont {Kurian}}, \ and\ \bibinfo {author}
  {\bibfnamefont {J.}~\bibnamefont {Prakash}},\ }\href {\doibase
  10.1103/PhysRevD.109.L011503} {\bibfield  {journal} {\bibinfo  {journal}
  {Phys. Rev. D}\ }\textbf {\bibinfo {volume} {109}},\ \bibinfo {pages}
  {L011503} (\bibinfo {year} {2024})},\ \Eprint
  {http://arxiv.org/abs/2311.16005} {arXiv:2311.16005 [hep-ph]} \BibitemShut
  {NoStop}%
\bibitem [{\citenamefont {Prakash}\ and\ \citenamefont
  {Jamal}(2024)}]{Prakash:2023zeu}%
  \BibitemOpen
  \bibfield  {author} {\bibinfo {author} {\bibfnamefont {J.}~\bibnamefont
  {Prakash}}\ and\ \bibinfo {author} {\bibfnamefont {M.~Y.}\ \bibnamefont
  {Jamal}},\ }\href {\doibase 10.1088/1361-6471/ad10c9} {\bibfield  {journal}
  {\bibinfo  {journal} {J. Phys. G}\ }\textbf {\bibinfo {volume} {51}},\
  \bibinfo {pages} {025101} (\bibinfo {year} {2024})}\BibitemShut {NoStop}%
\bibitem [{\citenamefont {Svetitsky}(1988)}]{Svetitsky:1987gq}%
  \BibitemOpen
  \bibfield  {author} {\bibinfo {author} {\bibfnamefont {B.}~\bibnamefont
  {Svetitsky}},\ }\href {\doibase 10.1103/PhysRevD.37.2484} {\bibfield
  {journal} {\bibinfo  {journal} {Phys. Rev. D}\ }\textbf {\bibinfo {volume}
  {37}},\ \bibinfo {pages} {2484} (\bibinfo {year} {1988})}\BibitemShut
  {NoStop}%
\bibitem [{\citenamefont {Walton}\ and\ \citenamefont
  {Rafelski}(2000)}]{Walton:1999dy}%
  \BibitemOpen
  \bibfield  {author} {\bibinfo {author} {\bibfnamefont {D.~B.}\ \bibnamefont
  {Walton}}\ and\ \bibinfo {author} {\bibfnamefont {J.}~\bibnamefont
  {Rafelski}},\ }\href {\doibase 10.1103/PhysRevLett.84.31} {\bibfield
  {journal} {\bibinfo  {journal} {Phys. Rev. Lett.}\ }\textbf {\bibinfo
  {volume} {84}},\ \bibinfo {pages} {31} (\bibinfo {year} {2000})},\ \Eprint
  {http://arxiv.org/abs/hep-ph/9907273} {arXiv:hep-ph/9907273} \BibitemShut
  {NoStop}%
\bibitem [{\citenamefont {Moore}\ and\ \citenamefont
  {Teaney}(2005)}]{Moore:2004tg}%
  \BibitemOpen
  \bibfield  {author} {\bibinfo {author} {\bibfnamefont {G.~D.}\ \bibnamefont
  {Moore}}\ and\ \bibinfo {author} {\bibfnamefont {D.}~\bibnamefont {Teaney}},\
  }\href {\doibase 10.1103/PhysRevC.71.064904} {\bibfield  {journal} {\bibinfo
  {journal} {Phys. Rev. C}\ }\textbf {\bibinfo {volume} {71}},\ \bibinfo
  {pages} {064904} (\bibinfo {year} {2005})},\ \Eprint
  {http://arxiv.org/abs/hep-ph/0412346} {arXiv:hep-ph/0412346} \BibitemShut
  {NoStop}%
\bibitem [{\citenamefont {Mazumder}\ \emph {et~al.}(2014)\citenamefont
  {Mazumder}, \citenamefont {Bhattacharyya},\ and\ \citenamefont
  {Alam}}]{Mazumder:2013oaa}%
  \BibitemOpen
  \bibfield  {author} {\bibinfo {author} {\bibfnamefont {S.}~\bibnamefont
  {Mazumder}}, \bibinfo {author} {\bibfnamefont {T.}~\bibnamefont
  {Bhattacharyya}}, \ and\ \bibinfo {author} {\bibfnamefont {J.-e.}\
  \bibnamefont {Alam}},\ }\href {\doibase 10.1103/PhysRevD.89.014002}
  {\bibfield  {journal} {\bibinfo  {journal} {Phys. Rev. D}\ }\textbf {\bibinfo
  {volume} {89}},\ \bibinfo {pages} {014002} (\bibinfo {year} {2014})},\
  \Eprint {http://arxiv.org/abs/1305.6445} {arXiv:1305.6445 [nucl-th]}
  \BibitemShut {NoStop}%
\bibitem [{\citenamefont {Liu}\ and\ \citenamefont {Rapp}(2020)}]{Liu:2020dlt}%
  \BibitemOpen
  \bibfield  {author} {\bibinfo {author} {\bibfnamefont {S.~Y.~F.}\
  \bibnamefont {Liu}}\ and\ \bibinfo {author} {\bibfnamefont {R.}~\bibnamefont
  {Rapp}},\ }\href {\doibase 10.1007/JHEP08(2020)168} {\bibfield  {journal}
  {\bibinfo  {journal} {JHEP}\ }\textbf {\bibinfo {volume} {08}},\ \bibinfo
  {pages} {168} (\bibinfo {year} {2020})},\ \Eprint
  {http://arxiv.org/abs/2003.12536} {arXiv:2003.12536 [nucl-th]} \BibitemShut
  {NoStop}%
\bibitem [{\citenamefont {Song}\ \emph {et~al.}(2023)\citenamefont {Song},
  \citenamefont {Grishmanovskii},\ and\ \citenamefont
  {Soloveva}}]{Song:2022wil}%
  \BibitemOpen
  \bibfield  {author} {\bibinfo {author} {\bibfnamefont {T.}~\bibnamefont
  {Song}}, \bibinfo {author} {\bibfnamefont {I.}~\bibnamefont
  {Grishmanovskii}}, \ and\ \bibinfo {author} {\bibfnamefont {O.}~\bibnamefont
  {Soloveva}},\ }\href {\doibase 10.1103/PhysRevD.107.036009} {\bibfield
  {journal} {\bibinfo  {journal} {Phys. Rev. D}\ }\textbf {\bibinfo {volume}
  {107}},\ \bibinfo {pages} {036009} (\bibinfo {year} {2023})},\ \Eprint
  {http://arxiv.org/abs/2210.04010} {arXiv:2210.04010 [nucl-th]} \BibitemShut
  {NoStop}%
\bibitem [{\citenamefont {Gyulassy}\ and\ \citenamefont
  {Wang}(1994)}]{Gyulassy:1993hr}%
  \BibitemOpen
  \bibfield  {author} {\bibinfo {author} {\bibfnamefont {M.}~\bibnamefont
  {Gyulassy}}\ and\ \bibinfo {author} {\bibfnamefont {X.-n.}\ \bibnamefont
  {Wang}},\ }\href {\doibase 10.1016/0550-3213(94)90079-5} {\bibfield
  {journal} {\bibinfo  {journal} {Nucl. Phys. B}\ }\textbf {\bibinfo {volume}
  {420}},\ \bibinfo {pages} {583} (\bibinfo {year} {1994})},\ \Eprint
  {http://arxiv.org/abs/nucl-th/9306003} {arXiv:nucl-th/9306003} \BibitemShut
  {NoStop}%
\bibitem [{\citenamefont {Klein}(1999)}]{Klein:1998du}%
  \BibitemOpen
  \bibfield  {author} {\bibinfo {author} {\bibfnamefont {S.}~\bibnamefont
  {Klein}},\ }\href {\doibase 10.1103/RevModPhys.71.1501} {\bibfield  {journal}
  {\bibinfo  {journal} {Rev. Mod. Phys.}\ }\textbf {\bibinfo {volume} {71}},\
  \bibinfo {pages} {1501} (\bibinfo {year} {1999})},\ \Eprint
  {http://arxiv.org/abs/hep-ph/9802442} {arXiv:hep-ph/9802442} \BibitemShut
  {NoStop}%
\bibitem [{\citenamefont {Mazumder}\ \emph {et~al.}(2015)\citenamefont
  {Mazumder}, \citenamefont {Bhattacharyya},\ and\ \citenamefont
  {Alam}}]{Mazumder:2015wma}%
  \BibitemOpen
  \bibfield  {author} {\bibinfo {author} {\bibfnamefont {S.}~\bibnamefont
  {Mazumder}}, \bibinfo {author} {\bibfnamefont {T.}~\bibnamefont
  {Bhattacharyya}}, \ and\ \bibinfo {author} {\bibfnamefont {J.-E.}\
  \bibnamefont {Alam}},\ }\href {\doibase 10.16943/ptinsa/2015/v81i1/48073}
  {\bibfield  {journal} {\bibinfo  {journal} {Proc. Indian Natl. Sci. Acad.}\
  }\textbf {\bibinfo {volume} {81}},\ \bibinfo {pages} {223} (\bibinfo {year}
  {2015})}\BibitemShut {NoStop}%
\bibitem [{\citenamefont {Abir}\ \emph {et~al.}(2012)\citenamefont {Abir},
  \citenamefont {Greiner}, \citenamefont {Martinez}, \citenamefont {Mustafa},\
  and\ \citenamefont {Uphoff}}]{Abir:2011jb}%
  \BibitemOpen
  \bibfield  {author} {\bibinfo {author} {\bibfnamefont {R.}~\bibnamefont
  {Abir}}, \bibinfo {author} {\bibfnamefont {C.}~\bibnamefont {Greiner}},
  \bibinfo {author} {\bibfnamefont {M.}~\bibnamefont {Martinez}}, \bibinfo
  {author} {\bibfnamefont {M.~G.}\ \bibnamefont {Mustafa}}, \ and\ \bibinfo
  {author} {\bibfnamefont {J.}~\bibnamefont {Uphoff}},\ }\href {\doibase
  10.1103/PhysRevD.85.054012} {\bibfield  {journal} {\bibinfo  {journal} {Phys.
  Rev. D}\ }\textbf {\bibinfo {volume} {85}},\ \bibinfo {pages} {054012}
  (\bibinfo {year} {2012})},\ \Eprint {http://arxiv.org/abs/1109.5539}
  {arXiv:1109.5539 [hep-ph]} \BibitemShut {NoStop}%
\bibitem [{\citenamefont {Kaczmarek}\ and\ \citenamefont
  {Zantow}(2005)}]{PhysRevD.71.114510}%
  \BibitemOpen
  \bibfield  {author} {\bibinfo {author} {\bibfnamefont {O.}~\bibnamefont
  {Kaczmarek}}\ and\ \bibinfo {author} {\bibfnamefont {F.}~\bibnamefont
  {Zantow}},\ }\href {\doibase 10.1103/PhysRevD.71.114510} {\bibfield
  {journal} {\bibinfo  {journal} {Phys. Rev. D}\ }\textbf {\bibinfo {volume}
  {71}},\ \bibinfo {pages} {114510} (\bibinfo {year} {2005})}\BibitemShut
  {NoStop}%
\bibitem [{\citenamefont {Acharya}\ \emph {et~al.}(2022)\citenamefont {Acharya}
  \emph {et~al.}}]{ALICE:2021aqk}%
  \BibitemOpen
  \bibfield  {author} {\bibinfo {author} {\bibfnamefont {S.}~\bibnamefont
  {Acharya}} \emph {et~al.} (\bibinfo {collaboration} {ALICE}),\ }\href
  {\doibase 10.1038/s41586-022-04572-w} {\bibfield  {journal} {\bibinfo
  {journal} {Nature}\ }\textbf {\bibinfo {volume} {605}},\ \bibinfo {pages}
  {440} (\bibinfo {year} {2022})},\ \bibinfo {note} {[Erratum: Nature 607, E22
  (2022)]},\ \Eprint {http://arxiv.org/abs/2106.05713} {arXiv:2106.05713
  [nucl-ex]} \BibitemShut {NoStop}%
\bibitem [{\citenamefont {Dokshitzer}\ and\ \citenamefont
  {Kharzeev}(2001)}]{Dokshitzer:2001zm}%
  \BibitemOpen
  \bibfield  {author} {\bibinfo {author} {\bibfnamefont {Y.~L.}\ \bibnamefont
  {Dokshitzer}}\ and\ \bibinfo {author} {\bibfnamefont {D.~E.}\ \bibnamefont
  {Kharzeev}},\ }\href {\doibase 10.1016/S0370-2693(01)01130-3} {\bibfield
  {journal} {\bibinfo  {journal} {Phys. Lett. B}\ }\textbf {\bibinfo {volume}
  {519}},\ \bibinfo {pages} {199} (\bibinfo {year} {2001})},\ \Eprint
  {http://arxiv.org/abs/hep-ph/0106202} {arXiv:hep-ph/0106202} \BibitemShut
  {NoStop}%
\bibitem [{\citenamefont {Das}\ \emph {et~al.}(2010)\citenamefont {Das},
  \citenamefont {Alam},\ and\ \citenamefont {Mohanty}}]{Das:2010tj}%
  \BibitemOpen
  \bibfield  {author} {\bibinfo {author} {\bibfnamefont {S.~K.}\ \bibnamefont
  {Das}}, \bibinfo {author} {\bibfnamefont {J.-e.}\ \bibnamefont {Alam}}, \
  and\ \bibinfo {author} {\bibfnamefont {P.}~\bibnamefont {Mohanty}},\ }\href
  {\doibase 10.1103/PhysRevC.82.014908} {\bibfield  {journal} {\bibinfo
  {journal} {Phys. Rev. C}\ }\textbf {\bibinfo {volume} {82}},\ \bibinfo
  {pages} {014908} (\bibinfo {year} {2010})},\ \Eprint
  {http://arxiv.org/abs/1003.5508} {arXiv:1003.5508 [nucl-th]} \BibitemShut
  {NoStop}%
\bibitem [{\citenamefont {Wong}(1970)}]{Wong:1970fu}%
  \BibitemOpen
  \bibfield  {author} {\bibinfo {author} {\bibfnamefont {S.~K.}\ \bibnamefont
  {Wong}},\ }\href {\doibase 10.1007/BF02892134} {\bibfield  {journal}
  {\bibinfo  {journal} {Nuovo Cim. A}\ }\textbf {\bibinfo {volume} {65}},\
  \bibinfo {pages} {689} (\bibinfo {year} {1970})}\BibitemShut {NoStop}%
\bibitem [{\citenamefont {Prakash}\ and\ \citenamefont
  {Jamal}(2023)}]{10.1088/1361-6471/ad10c9}%
  \BibitemOpen
  \bibfield  {author} {\bibinfo {author} {\bibfnamefont {J.}~\bibnamefont
  {Prakash}}\ and\ \bibinfo {author} {\bibfnamefont {M.~Y.}\ \bibnamefont
  {Jamal}},\ }\href
  {http://iopscience.iop.org/article/10.1088/1361-6471/ad10c9} {\bibfield
  {journal} {\bibinfo  {journal} {Journal of Physics G: Nuclear and Particle
  Physics}\ } (\bibinfo {year} {2023})}\BibitemShut {NoStop}%
\bibitem [{\citenamefont {Jamal}\ \emph {et~al.}(2017)\citenamefont {Jamal},
  \citenamefont {Mitra},\ and\ \citenamefont {Chandra}}]{Jamal:2017dqs}%
  \BibitemOpen
  \bibfield  {author} {\bibinfo {author} {\bibfnamefont {M.~Y.}\ \bibnamefont
  {Jamal}}, \bibinfo {author} {\bibfnamefont {S.}~\bibnamefont {Mitra}}, \ and\
  \bibinfo {author} {\bibfnamefont {V.}~\bibnamefont {Chandra}},\ }\href
  {\doibase 10.1103/PhysRevD.95.094022} {\bibfield  {journal} {\bibinfo
  {journal} {Phys. Rev. D}\ }\textbf {\bibinfo {volume} {95}},\ \bibinfo
  {pages} {094022} (\bibinfo {year} {2017})},\ \Eprint
  {http://arxiv.org/abs/1701.06162} {arXiv:1701.06162 [nucl-th]} \BibitemShut
  {NoStop}%
\bibitem [{\citenamefont {Kumar}\ \emph {et~al.}(2018)\citenamefont {Kumar},
  \citenamefont {Jamal}, \citenamefont {Chandra},\ and\ \citenamefont
  {Bhatt}}]{Kumar:2017bja}%
  \BibitemOpen
  \bibfield  {author} {\bibinfo {author} {\bibfnamefont {A.}~\bibnamefont
  {Kumar}}, \bibinfo {author} {\bibfnamefont {M.~Y.}\ \bibnamefont {Jamal}},
  \bibinfo {author} {\bibfnamefont {V.}~\bibnamefont {Chandra}}, \ and\
  \bibinfo {author} {\bibfnamefont {J.~R.}\ \bibnamefont {Bhatt}},\ }\href
  {\doibase 10.1103/PhysRevD.97.034007} {\bibfield  {journal} {\bibinfo
  {journal} {Phys. Rev. D}\ }\textbf {\bibinfo {volume} {97}},\ \bibinfo
  {pages} {034007} (\bibinfo {year} {2018})},\ \Eprint
  {http://arxiv.org/abs/1709.01032} {arXiv:1709.01032 [nucl-th]} \BibitemShut
  {NoStop}%
\bibitem [{\citenamefont {Cacciari}\ \emph {et~al.}(2005)\citenamefont
  {Cacciari}, \citenamefont {Nason},\ and\ \citenamefont
  {Vogt}}]{Cacciari:2005rk}%
  \BibitemOpen
  \bibfield  {author} {\bibinfo {author} {\bibfnamefont {M.}~\bibnamefont
  {Cacciari}}, \bibinfo {author} {\bibfnamefont {P.}~\bibnamefont {Nason}}, \
  and\ \bibinfo {author} {\bibfnamefont {R.}~\bibnamefont {Vogt}},\ }\href
  {\doibase 10.1103/PhysRevLett.95.122001} {\bibfield  {journal} {\bibinfo
  {journal} {Phys. Rev. Lett.}\ }\textbf {\bibinfo {volume} {95}},\ \bibinfo
  {pages} {122001} (\bibinfo {year} {2005})},\ \Eprint
  {http://arxiv.org/abs/hep-ph/0502203} {arXiv:hep-ph/0502203} \BibitemShut
  {NoStop}%
\bibitem [{\citenamefont {Cacciari}\ \emph {et~al.}(2012)\citenamefont
  {Cacciari}, \citenamefont {Frixione}, \citenamefont {Houdeau}, \citenamefont
  {Mangano}, \citenamefont {Nason},\ and\ \citenamefont
  {Ridolfi}}]{Cacciari:2012ny}%
  \BibitemOpen
  \bibfield  {author} {\bibinfo {author} {\bibfnamefont {M.}~\bibnamefont
  {Cacciari}}, \bibinfo {author} {\bibfnamefont {S.}~\bibnamefont {Frixione}},
  \bibinfo {author} {\bibfnamefont {N.}~\bibnamefont {Houdeau}}, \bibinfo
  {author} {\bibfnamefont {M.~L.}\ \bibnamefont {Mangano}}, \bibinfo {author}
  {\bibfnamefont {P.}~\bibnamefont {Nason}}, \ and\ \bibinfo {author}
  {\bibfnamefont {G.}~\bibnamefont {Ridolfi}},\ }\href {\doibase
  10.1007/JHEP10(2012)137} {\bibfield  {journal} {\bibinfo  {journal} {JHEP}\
  }\textbf {\bibinfo {volume} {10}},\ \bibinfo {pages} {137} (\bibinfo {year}
  {2012})},\ \Eprint {http://arxiv.org/abs/1205.6344} {arXiv:1205.6344
  [hep-ph]} \BibitemShut {NoStop}%
\end{thebibliography}%
\end{document}